\documentclass{SciPost}

\hypersetup{
    colorlinks,
    linkcolor={red!50!black},
    citecolor={blue!50!black},
    urlcolor={blue!80!black}
}

\usepackage[bitstream-charter]{mathdesign}
\DeclareSymbolFont{usualmathcal}{OMS}{cmsy}{m}{n}
\DeclareSymbolFontAlphabet{\mathcal}{usualmathcal}

\fancypagestyle{SPstyle}{
\fancyhf{}
\lhead{\colorbox{scipostblue}{\bf \color{white} ~SciPost Physics Core }}
\rhead{{\bf \color{scipostdeepblue} ~Submission }}

\fancyfoot[C]{\textbf{\thepage}}
}
\usepackage[version=3]{mhchem}
\usepackage{braket}
\usepackage{gensymb}
\usepackage[T1]{fontenc}
\usepackage{soul}
\usepackage[symbol]{footmisc}
\usepackage{amsmath}
\usepackage{physics}
\usepackage{graphicx}
\usepackage{graphics}
\usepackage{hyperref}

\newcommand{\kdotp}{\ensuremath{\mathbf{k}\cdot\mathbf{p}}}

\begin{document}

\pagestyle{SPstyle}

\begin{center}{\Large \textbf{\color{scipostdeepblue}{
%%%%%%%%%% TODO: Write your article's title here
Anisotropic sub-band splitting mechanisms in strained HgTe: a first principles study\\
}}}\end{center}

\begin{center}\textbf{
% Use (full) first name (+ middle name initials) + surname format.
% Separate subsequent authors by a comma, omit comma and use "and" for the last author.
% Mark the corresponding author(s) with a superscript symbol in this order
% \star, \dagger, \ddagger, \circ, \S, \P, \parallel, ...
Eeshan Ketkar\textsuperscript{1,2,3$\star$},
Giovanni Marini\textsuperscript{4},
Pietro Maria Forcella\textsuperscript{5},
Giorgio Sangiovanni\textsuperscript{6},
Gianni Profeta\textsuperscript{5,7}, and
Wouter Beugeling\textsuperscript{1,2}
}\end{center}

\begin{center}
{\bf 1} Physikalisches Institut (EP3), Universit\"{a}t W\"{u}rzburg, Am Hubland, 97074 W\"{u}rzburg, Germany
\\
{\bf 2} Institute for Topological Insulators, Am Hubland, 97074 W\"{u}rzburg, Germany
\\
{\bf 3} Pritzker School of Molecular Engineering, The University of Chicago, Chicago, Illinois 60637, USA
\\
{\bf 4} {Department of Physics, University of Trento, Via Sommarive 14, 38123 Povo}, {Italy}
\\
{\bf 5} Dipartimento di Scienze Fisiche e Chimiche, Università degli Studi dell’Aquila, Via Vetoio 10, I-67100 L'Aquila, Italy
\\
{\bf 6} {Institut f{\"u}r Theoretische Physik und Astrophysik and W{\"u}rzburg-Dresden Cluster of Excellence ct.qmat, Universit{\"a}t W{\"u}rzburg, 97074 W{\"u}rzburg}, {Germany}
\\
{\bf 7} CNR-SPIN c/o Dipartimento di Scienze Fisiche e Chimiche, Università degli Studi dell’Aquila, Via Vetoio 10, I-67100 L’Aquila, Italy

%%%%%%%%%% END TODO: AFFILIATIONS
% Provide email address of corresponding author(s)
$\star$ \href{mailto:email1}{\small eketkar@uchicago.edu}
\end{center}

\section*{\color{scipostdeepblue}{Abstract}}
\textbf{\boldmath{%
Mercury telluride is a canonical material for realizing topological phases, yet a full understanding of its electronic structure remains challenging due to subtle competing effects. Using first-principles calculations and \kdotp{} modelling, we study its topological phase diagram under strain. We show that linearly $k$-dependent higher-order $C_4$ strain terms are important for capturing the correct low-energy behaviour. These terms lead to a nontrivial $k$-dependence of the sub-band splitting arising from the interplay of strain and bulk inversion asymmetry. This explains the camel-back feature in the tensile regime and supports the emergence of a Weyl semimetal phase under compressive strain.
}}

\vspace{\baselineskip}

%%%%%%%%%% BLOCK: Copyright information
% This block will be filled during the proof stage, and finilized just before publication.
% It exists here only as a placeholder, and should not be modified by authors.
\noindent\textcolor{white!90!black}{%
\fbox{\parbox{0.975\linewidth}{%
\textcolor{white!40!black}{\begin{tabular}{lr}%
  \begin{minipage}{0.6\textwidth}%
    {\small Copyright attribution to authors. \newline
    This work is a submission to SciPost Physics Core. \newline
    License information to appear upon publication. \newline
    Publication information to appear upon publication.}
  \end{minipage} & \begin{minipage}{0.4\textwidth}
    {\small Received Date \newline Accepted Date \newline Published Date}%
  \end{minipage}
\end{tabular}}
}}
}
%%%%%%%%%% BLOCK: Copyright information

% For convenience during refereeing we turn on line numbers:
%\linenumbers

\vspace{10pt}
\noindent\rule{\textwidth}{1pt}
\tableofcontents
\noindent\rule{\textwidth}{1pt}
\vspace{10pt}

% Write your article contents here, starting from first \section.
% An example structure is given below.

\section{Introduction}
\label{sec:intro}
% TODO: write your article here.
Topological properties of solid-state systems have attracted large interest in the last years. New states of matter have been observed and a new paradigm has been introduced to describe phase transitions that cannot be characterized within the classical Landau theory \cite{PhysRevLett.95.146802,PhysRevLett.98.106803,RevModPhys.83.1057,PhysRevB.75.121306,RevModPhys.82.3045}. Mercury telluride (\ce{HgTe}) has played a crucial role in this regard, being the first platform where many of these ideas found experimental realization \cite{doi:10.1126/science.1148047}. 
\par
The nontrivial topological properties of \ce{HgTe} are related to its low-energy electronic structure around the $\Gamma$ point of the Brillouin zone. In the unstrained state, \ce{HgTe} is semimetallic in nature whereas a band inversion between Hg(s) and Te(p) states can be observed when a considerable tensile strain is applied on the system \cite{PhysRevLett.117.086403,Leubner2017}. Under compressive strain and much smaller tensile strain, \ce{HgTe} and its superlattices transform into a Weyl semimetal \cite{ruan2016symmetry,chen2019strain,PhysRevX.9.031034,AravindnathEA2025,PhysRevB.87.045202,PhysRevLett.132.016603,PhysRevResearch.4.023114,kharitonov2022evolution}. 
\par
The advanced \kdotp{} models typically employed to describe the electronic band structure \cite{Kane1957,ruan2016symmetry,PhysRevB.72.035321,chen2019strain} exhibit important quantitative differences, such as the magnitude of sub-band splitting along specific $k$-paths, when compared to those predicted by first principles calculations. Conversely, these first principles calculations are in very good agreement with respect to the angle-resolved photoemission spectroscopy (ARPES) experiments \cite{PhysRevB.107.L121102}. While such energy splittings are relatively small, they are important for two reasons: The first is that the tensile-strained \ce{HgTe} gap is very small. The second is that the appearance of such splittings underlie the existence of other $k$-dependent terms, whose understanding may be crucial for materials design and for the inherent comprehension of \ce{HgTe} physics, including the camel's back formation in the tensile strain phase and the topological phase transition towards a Weyl phase as a function of strain \cite{ruan2016symmetry,chen2019strain}.
\par
In the present study we employ a perturbed 8 band \kdotp{} model fitted to state of the art density-functional theory (DFT) calculations \cite{PhysRevB.107.L121102}, able to quantitatively describe the photoemission spectra of \ce{HgTe}, in order to identify the underlying factors responsible for the band splitting along multiple crystallographic directions. We find that the band splitting along a particular crystallographic direction arises from a competition between the first-order strain in the momentum coordinate $k$ perturbation term ($C_4$) \cite{PhysRevB.20.686} and the bulk-inversion asymmetry (BIA) term, which stems from the non-centrosymmetric nature of the \ce{HgTe} lattice \cite{PhysRev.100.580,semenikhin2007effects}.
The $C_4$ strain terms were neglected by previous models \cite{ruan2016symmetry,PhysRevB.72.035321,Leubner2017,mahler2021massive,zhang2001rashba,becker2000band,zhang2004effective,chen2019strain}, resulting in the absence of $k$-dependent strain-induced sub-band splitting. Here, we establish the necessity of incorporating these $k$-dependent $C_4$ strain terms into the $8\times 8$ Kane Hamiltonian \cite{Kane1957} to model the experimental electronic band structure. 
We find that band splitting is primarily induced by the $C_4$ strain terms for crystallographic directions in close proximity to or along the $k_x$, $k_y$ and $k_z$ axes. Such splittings were not captured by previous models, which only considered $k$ independent strain terms with BIA, and showed negligible splitting in proximity to these axes. We proceed to highlight the competition between the $C_4$ and BIA terms in the sub-band splitting mechanism and finally, we gauge the effect of these strain terms on the topological strain phase diagram of \ce{HgTe} \cite{ruan2016symmetry,chen2019strain}, demonstrating the robustness of the topological Weyl semimetal state with respect to them. Interestingly, we find that our model results in a tilted type-1 Weyl semimetal state instead of the ideal Weyl semimetal state observed in prior work \cite{ruan2016symmetry,chen2019strain}. Such a tilt of the Weyl cones enhances the Berry curvature dipole \cite{zhang2018berry} and can be used to explain the superconducting diode effect \cite{PhysRevB.109.064511}.
\begin{figure*}[t!]
\includegraphics[width=0.9\textwidth]{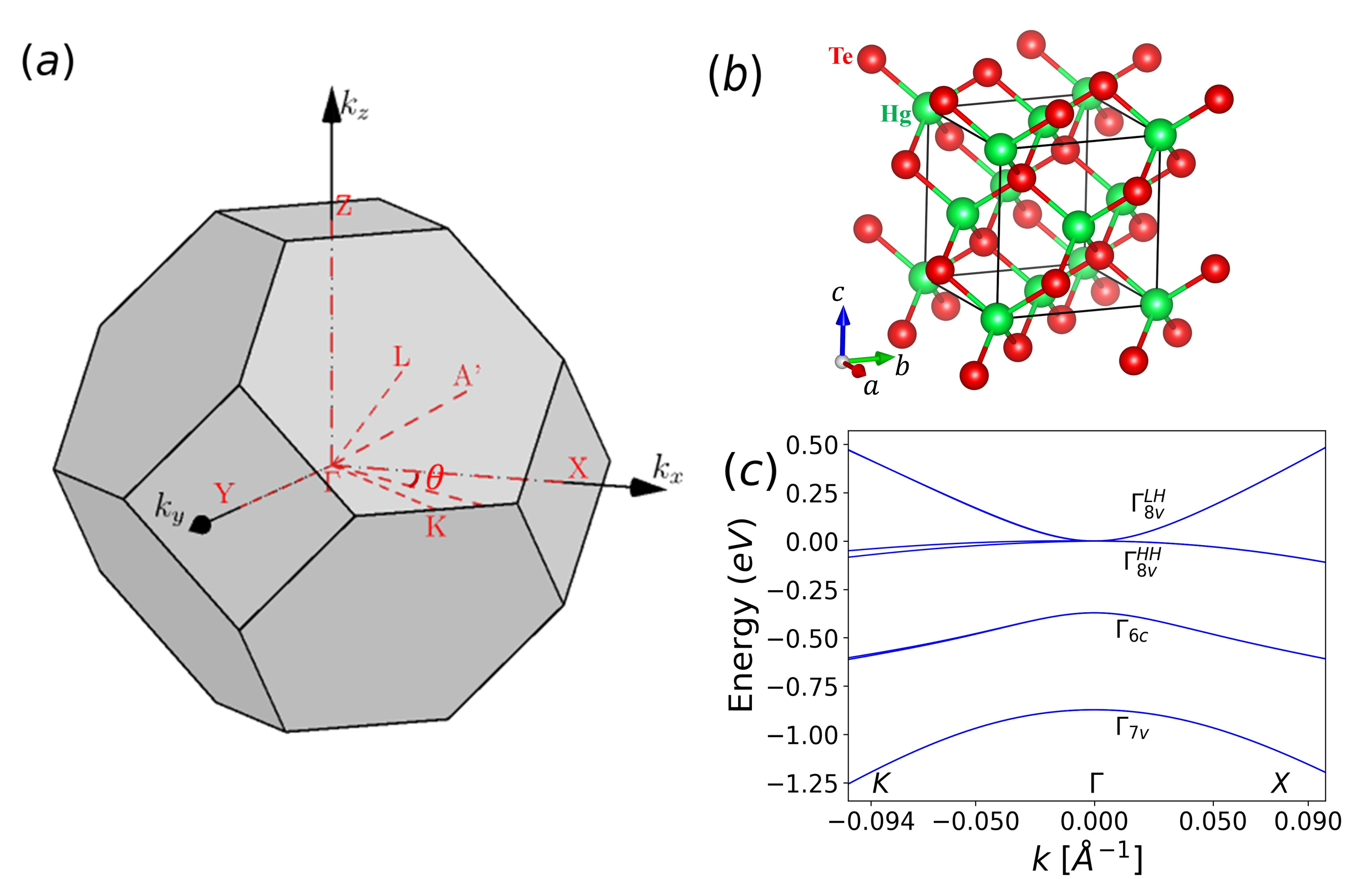}
    \caption{(a) The first Brillouin zone of the \ce{HgTe} lattice depicting the $k$-paths used in this study (b) The crystal structure of \ce{HgTe} (c) The electronic band structure of unstrained \ce{HgTe} calculated by first principles.}
 \label{fig.1}
\end{figure*}

This paper is structured as follows: in Sec.~\ref{cd} we provide the computational details, in Sec.~\ref{res} we present and discuss the results of our detailed analysis, and finally in Sec.~\ref{conc} we conclude. 

\section{Modelling}
\label{cd}
\subsection{Density-functional theory calculations}

We performed DFT calculations with the projector augmented-wave pseudopotential method \cite{PhysRevB.50.17953,PhysRevB.59.1758} as implemented in the Vienna Ab-initio Simulation Package (VASP) \cite{PhysRevB.47.558,PhysRevB.49.14251,KRESSE199615,PhysRevB.54.11169}. An energy cutoff of $350$ eV for the plane wave basis and $8\times8\times8$ Monkhorst-Pack grid for Brillouin zone sampling were used, ensuring a convergence of 1 meV on the electronic eigenvalues. Spin-orbit coupling was included in the calculation. To simulate tensile-strained \ce{HgTe}, we considered the lattice parameter $a_0 = 6.46$ \AA{} for \ce{HgTe} in its pristine state, and applied a $0.31\%$ in-plane tensile strain in order to match the \ce{CdTe} lattice parameter ($0.31\%$ tensile biaxial strain, $a_f = 6.481$ \AA, where $a_f$ denotes the lattice constant post-deformation). A compressed-strained \ce{HgTe} was considered to simulate the Weyl semimetal phase ($0.5\%$ compressive biaxial strain, $a_f = 6.429$ \AA). The corresponding out of plane lattice parameter can be calculated from the stiffness coefficients of \ce{HgTe} \cite{Leubner2017,Berding2000}. We obtain $c_f = 6.435$ \AA\, for the tensile biaxial strain, and $c_f = 6.525$ \AA{} for the compressive-strained phase. We employed the hybrid HSE06 functional \cite{Krukau2006}, explicitly including a fraction of the exact-exchange term. The choice for the exchange-correlation functional is justified by the comparative analysis performed in our previous work \cite{PhysRevB.107.L121102}, where the superior performance of HSE06 with respect to other functionals was attested. The search for Weyl points in the compressed phase of \ce{HgTe} were performed using the \texttt{WannierTools} code \cite{WU2017}, with a tight-binding Hamiltonian mapped from a DFT band structure using the \texttt{Wannier90} code \cite{Pizzi2020} on an $8 \times 8 \times 8$ $k$-point grid.

\subsection{\kdotp{} Theory}\label{k.p}

\begin{figure*}[t]
\includegraphics[width=0.95\textwidth]{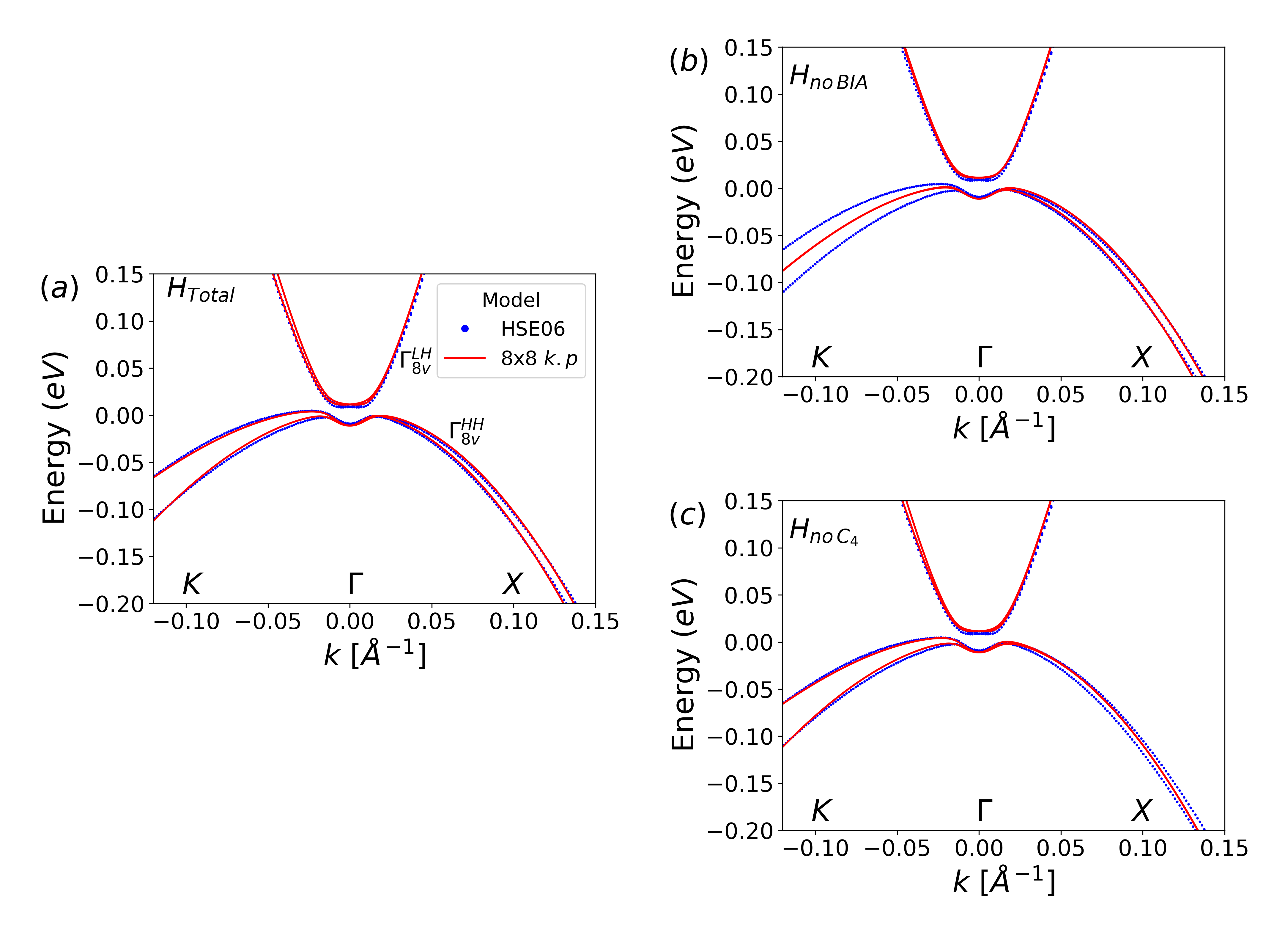}
    \caption{The \kdotp{} electronic band structure fit to the DFT results including (a) both BIA and $C_4$ strain terms ($H_{\mathrm{Total}}$) (b) only $C_{4}$ strain terms ($H_{\mathrm{no\,BIA}}$) and (c) only BIA terms ($H_{\mathrm{no\,C_{4}}}$). The $\Gamma_{8v}^{HH}$ bands correspond to the $\ket{\Gamma_{8}, \pm\frac{3}{2}}$ basis whereas the $\Gamma_{8v}^{LH}$ bands correspond to the $\ket{\Gamma_{8}, \pm\frac{1}{2}}$ basis.}
 \label{fig.2}
\end{figure*}

In order to better understand the underlying physics of our DFT calculations, we fit an $8\times8$ Kane \kdotp{} model Hamiltonian $H_{\mathrm{Kane}}$ for zincblende lattices \cite{Kane1957} to the DFT electronic band structure. Our \kdotp{} model Hamiltonian has been constructed from the 8 orbital basis set comprising of ${\ket{\Gamma_6,\pm\frac{1}{2}}}$, ${\ket{\Gamma_8,\pm\frac{1}{2}}}$, ${\ket{\Gamma_8,\pm\frac{3}{2}}}$ and ${\ket{\Gamma_7,\pm\frac{1}{2}}}$ as described in \cite{winkler2003spin} (see Appendix section \ref{hkanedef} for the matrix definition).
Since our \ce{HgTe} lattice has been subjected to axial tensile or compressive strain we describe its strain tensor $\begin{bmatrix} \epsilon_{ij} \end{bmatrix}$ as
\begin{equation}
\begin{bmatrix} \epsilon_{ij} \end{bmatrix} = 
\begin{bmatrix}
  \epsilon_{xx} & 0 & 0\\ 
  0 & \epsilon_{yy} & 0\\
  0 & 0 & \epsilon_{zz}
\end{bmatrix}, 
\end{equation}
where $\epsilon_{ii} = (a_f)_i / (a_0)_i - 1, i={x,y,z}$ in terms of the post-strain and equilibrium lattice constants, $(a_{f})_i$ and $(a_{0})_i$, respectively.
For our calculations in the topological insulator (TI) state we consider $\epsilon_{xx}$ = $\epsilon_{yy}$ = 0.31\% for \ce{HgTe} epitaxially strained to a \ce{CdTe} substrate. For the Weyl semimetal state, we consider $\epsilon_{xx}$ = $\epsilon_{yy} = -0.5\%$.
The ratio $\epsilon_{zz} / \epsilon_{xx} = -1.38$, which has been obtained from the elasticity coefficients \cite{alper1967elastic} ratio $C_{12}/C_{11}$ for epitaxial growth along the (001) direction, is used for the entirety of our calculations. The effect of strain on the \ce{HgTe} lattice is accounted for by the strain Hamiltonian proposed by Pikus and Bir \cite{bir1974symmetry} $H_{\mathrm{Pikus\text{--}Bir}}$ as described in Ref. \cite{PhysRevB.72.035321} (see Appendix section \ref{hpikusbirdef} for the matrix definition).
Prior \kdotp{} models \cite{ruan2016symmetry,PhysRevB.72.035321,Leubner2017,mahler2021massive,zhang2001rashba,becker2000band,zhang2004effective,chen2019strain} used for the electronic band structure analysis of strained \ce{HgTe} have been unable to explain the root cause of sub-band splitting. Our work successfully solves this long standing problem by considering the $k$-dependent $C_{4}$ strain terms \cite{PhysRevB.20.686}  obtained through perturbation theory, in addition to the Pikus-Bir strain terms \cite{winkler2003spin}. The Hamiltonian matrix $H_{C_4}$ can be represented as
\begin{equation}\label{eq:hc4}
H_{C_{4}} = 
\begin{bmatrix}
  0 & 0 & 0\\ 
  0 & H_{C_{4}}^{8v,8v} & H_{C_{4}}^{8v,7v}\\
  0 & (H_{C_{4}}^{8v,7v})^{\dag} & 0
\end{bmatrix},
\end{equation}
where $H_{C_{4}}^{8v,8v}$ represents the interactions between the $\Gamma_{8v}$ bands which can be described as
\begin{equation}\label{defC488}
H_{C_{4}}^{8v,8v} = C_{4}\left[(\epsilon_{yy} -\epsilon_{zz})k_{x}J_{x}
+ (\epsilon_{zz} -\epsilon_{xx})k_{y}J_{y} + (\epsilon_{xx} - \epsilon_{yy})k_{z}J_{z}\right],
\end{equation}
where the $J_{a}$ matrices, $a = (x, y, z)$ represent angular momentum matrices (see Appendix section \ref{JUdef} for the definition).
$H_{C_{4}}^{8v,7v}$ represents the interactions between the $\Gamma_{8v}$ and $\Gamma_{7v}$ bands which can be described as 
\begin{equation}\label{defC487}
H_{C_{4}}^{8v,7v} = \frac{3}{2}C_{4}\left[(\epsilon_{yy} - \epsilon_{zz})k_{x}U_{x}
+ (\epsilon_{zz} - \epsilon_{xx})k_{y}U_{y} + (\epsilon_{xx} - \epsilon_{yy})k_{z}U_{z}\right],
\end{equation}
where the $U_{a}$ matrices, $a = (x, y, z)$ are representative of the interactions between the $\Gamma_{8v}$ bands and the $\Gamma_{7v}$ bands (see Appendix section \ref{JUdef} for the definitions).

\begin{table}[b]
\resizebox{\columnwidth}{!}{%
\begin{tabular}{|*{6}{c|}}
\hline
Path           & $\gamma_1$ & $\gamma_2$ & $\gamma_3$ & $C$ ($\mathrm{eV \mathring{A}}$) & $C_4$ ($\mathrm{eV \mathring{A}}$)            \\ \hline
$K$-$\Gamma$-$X$ \& $Z$-$\Gamma$-$X$ & 3.802  & 0.385 & 1.220  & 0.113 & 6.400\\ \hline
$\theta$-$\Gamma$-$X$, $\theta = 15\degree$ & 3.803  & 0.386 & 1.270  & 0.123 & 6.400\\ \hline
$L$-$\Gamma$-$X$          & 3.805  & 0.384 & 1.290 & 0.131 & 6.400\\ \hline
$A'$-$\Gamma$-$X$          & 3.807  & 0.383 & 1.000 & 0.119 & 6.400\\ \hline
Mean fit parameters (TI state)         & 3.804  & 0.385 & 1.195 & 0.126 & 6.400\\ \hline
$K$-$\Gamma$-$X$ (Weyl semimetal state)         & 4.100  & 0.570 & 1.23 & 0.131 & 6.400\\ \hline
\end{tabular}%
}
\caption{\label{table1} The fit parameters obtained on fitting $H_{\mathrm{Total}}$ to DFT HSE06 band structure data.}
\end{table}

The non-centrosymmetric nature of the \ce{HgTe} lattice results in the absence of inversion symmetry \cite{PhysRev.100.580,semenikhin2007effects}. To account for this, we also include the bulk inversion asymmetry (BIA) matrix $H_{\mathrm{BIA}}$ in our calculations with terms described in Ref.~\cite{winkler2003spin} (see Appendix section \ref{biadef}). Thus, the Hamiltonian used for our fit to DFT data can be expressed as 
%
%\begin{equation}\label{netHam}
%H_{\mathrm{Total}} = H_{\mathrm{Kane}} + H_{\mathrm{Pikus\text{--}Bir}} + H_{\mathrm{BIA}} + H_{C_{4}}
%\end{equation}
\begin{equation}
H_{\mathrm{Total}}(\mathbf{k},\varepsilon)=H_{\mathrm{Kane}}(\mathbf{k})
+H_{\mathrm{Pikus\text{--}Bir}}(\varepsilon)
+H_{\mathrm{BIA}}(\mathbf{k})
+H_{C_4}(\mathbf{k},\varepsilon),
\label{eq:HTotal}
\end{equation}
with the strain tensor $\varepsilon=\mathrm{diag}(\varepsilon_\parallel,\varepsilon_\parallel,\varepsilon_\perp)$ being diagonal, with $\varepsilon_{xx}=\varepsilon_{yy}\equiv \varepsilon_\parallel$ and $\varepsilon_{zz}\equiv \varepsilon_\perp$, corresponding to biaxial epitaxial strain along $(001)$.

Since the $C_{4}$ strain terms and BIA terms represent a very small deformation to the sum of the $8\times 8$ Kane Hamiltonian and the standard Pikus-Bir strain terms, we treat them perturbatively (see section \ref{origin} for the definition).
In order to gauge the robustness of the coefficients affiliated with our \kdotp{} model, we fit $H_{\mathrm{Total}}$ to our DFT band structure along different paths in the Brillouin zone using least squares regression.
Since leaving all the fitting parameters free leads to an uncontrolled result, which does not converge, we only consider the Luttinger coefficients ($\gamma_1$, $\gamma_2$ and $\gamma_3$, see Appendix section \ref{hkanedef} for more details), the linear BIA term $C$ (see Appendix section \ref{biadef}), and the $C_4$ strain term ,because these have a profound effect on band splitting and curvature. The values of these parameters obtained through fitting for the TI state are listed in Table \ref{table1}.
The obtained values of $\gamma_1$, $\gamma_2$ and $C_4$ are highly robust and show negligible variation with different paths along the Brillouin zone, while $\gamma_3$ and $C$ vary slightly by about $\pm10$\% and $\pm20$\%, respectively. Owing to the complexity of fitting our \kdotp{} model across the entire 3D Brillouin zone, we chose to repeat the least squares regression fit for different 1D paths to obtain a more accurate set of parameters, specific to those particular paths.

\section{Results and discussion}
\label{res}

\subsection{Origin of band splitting in strained \ce{HgTe}}\label{origin}

\begin{table}[bt]
\centering
\caption{Summary of point-group symmetries associated with different Hamiltonian contributions.}
\label{tab:pointGroupSummary}

\begin{tabular}{ll}
\hline
\textbf{Hamiltonian} & \textbf{Point Group Symmetry} \\
\hline
$H_{\mathrm{Kane}}$ & $O_h$ \\
$H_{\mathrm{Kane}} + H_{\mathrm{Pikus\text{--}Bir}}$ & $D_{4h}$ \\
$H_{\mathrm{Kane}} + H_{\mathrm{BIA}}$ & $T_d$ \\
$H_{\mathrm{Kane}} + H_{\mathrm{Pikus\text{--}Bir}} + H_{\mathrm{BIA}}$ & $D_{2d}$ \\
$H_{\mathrm{Kane}} + H_{\mathrm{Pikus\text{--}Bir}} + H_{C_4}$ & $D_{2d}$ \\
$H_{\mathrm{Total}} = H_{\mathrm{Kane}} + H_{\mathrm{Pikus\text{--}Bir}} + H_{\mathrm{BIA}} + H_{C_4}$ & $D_{2d}$ \\
\hline
\end{tabular}
\end{table}

For a better quantitative analysis of the effect of strain and BIA symmetry breaking terms on the sub-band splitting in our \ce{HgTe} system, we fit a perturbed $8\times 8$ Kane Hamiltonian (Eq.~\eqref{eq:HTotal}) to the DFT band structure along the $K$-$\Gamma$-$X$ path (Fig. \ref{fig.2}(a)).
In order to separate the effects of the $C_{4}$ strain term from those of the BIA terms, we will consider variations of Eq.~\eqref{eq:HTotal}, where we ignore $H_{\mathrm{BIA}}$ or  $H_{C_4}$, i.e., $H_{\mathrm{no\,BIA}} = H_{\mathrm{Kane}} + H_{\mathrm{Pikus\text{--}Bir}} + H_{C_4}$ or $H_{\mathrm{Kane}} + H_{\mathrm{Pikus\text{--}Bir}} + H_{\mathrm{BIA}}$, respectively.
In Table~\ref{tab:pointGroupSummary}, we list the point-group symmetries of all relevant combinations of the terms in Eq.~\eqref{eq:HTotal}. A detailed symmetry analysis is provided in Appendix section~\ref{app:symmetry}.

We first study the effects of the $C_{4}$ strain by ignoring $H_{\mathrm{BIA}}$, i.e., by considering $H_{\mathrm{no\,BIA}}$.
To provide a clearer explanation of the band splitting observed in the $\Gamma_{8v}^{HH}$ heavy hole (HH) bands (i.e., the bands associated with ${\ket{\Gamma_8,\pm\frac{3}{2}}}$), we make the following assumptions to simplify the blocks of $H_{C_{4}}$ along the crystallographic directions of interest. In our \ce{HgTe} system, we consider the in-plane strain to be isotropic, hence $\epsilon_{xx} = \epsilon_{yy}$ and $(\epsilon_{zz} - \epsilon_{xx}) = -(\epsilon_{yy} - \epsilon_{zz})$. Thus the expressions in Eqs.~\eqref{defC488} and \eqref{defC487} are simplified to
\begin{equation}\label{cancelZ88}
H_{C_{4}}^{8v,8v} = C_{4}(\epsilon_{yy} - \epsilon_{zz})(k_{x}J_{x} - k_{y}J_{y}),
\end{equation}
\begin{equation}\label{cancelZ87}
H_{C_{4}}^{8v,7v} = \frac{3}{2}C_{4}(\epsilon_{yy} - \epsilon_{zz})(k_{x}U_{x} - k_{y}U_{y}).
\end{equation}
In the absence of BIA and $C_{4}$ strain terms there is no splitting between the energy bands, and the eigenvalues of $H_{\mathrm{Kane}} + H_{\mathrm{Pikus\text{--}Bir}}$ are pairwise spin-degenerate. Therefore, to accurately gauge the effect of the $C_4$ strain terms on the band splitting between $\Gamma_{8v}^{HH}$ bands, we must apply the formalism of degenerate perturbation theory.
\par
Let $\ket{\Psi_1}$ and $\ket{\Psi_2}$ be an orthonormal basis constructed from the normalized eigenstates of $H_{\mathrm{Kane}} + H_{\mathrm{Pikus\text{--}Bir}}$ corresponding to the degenerate $\Gamma_{8v}^{HH}$ bands. On rewriting $H_{C_4}$ in terms of $\ket{\Psi_{n = 1,2}}$ and after diagonalization, we obtain the the eigenvalues $E_1$ and $E_2$, from which we can define a parameter: $\Delta E_{C_4} = E_2 - E_1$, which represents splitting between the $\Gamma_{8v}^{HH}$ bands. Calculating the value of $\Delta E_{C_{4}}$ at a point along a particular crystallographic direction gives us an idea of whether the splitting observed along that path is significant or negligible.
\par
Along the $\Gamma$-$X$ direction, $k_{y} = 0$, so Eqns. \eqref{cancelZ88} and \eqref{cancelZ87} can be written as 
\begin{equation}\label{eq:C488-gamma-x}
H_{C_{4}}^{8v,8v} = C_{4}(\epsilon_{yy} - \epsilon_{zz})k_{x}J_{x}, 
\end{equation}
\begin{equation}
H_{C_{4}}^{8v,7v} = \frac{3}{2}C_{4}(\epsilon_{yy} - \epsilon_{zz})k_{x}U_{x}.
\end{equation}
For example, at $k = 0.1$ $\mathrm{\mathring{A}^{-1}}$ we obtain the eigenvalues $E_1 = -7.18$ $\mathrm{meV}$ and $E_2 = 7.28$ $\mathrm{meV}$, which results in $\Delta E_{C_{4}} = 14.46$ $\mathrm{meV}$. Thus, the calculated value of $\Delta E_{C_{4}}$ along this direction is large enough to induce a noticeable splitting between the ${\Gamma_{8v}^{HH}}$. This implies that the $C_4$ strain terms will contribute significantly to band splitting along the $\Gamma$-$X$ direction.
Symmetry considerations (see Appendix section \ref{app:high-symmetry-lines}) show that terms linear in momentum that transform like axial vectors should vanish along $\Gamma$-$X$, so that the spin splitting from BIA is strongly suppressed. On the other hand, the term of Eq.~\eqref{eq:C488-gamma-x} induces band splitting by mixing bands within the valence band. As a result, $H_{C_4}$ dominates the observed splitting along $\Gamma$-$X$.

\begin{figure*}[t]
\includegraphics[width=1.02\textwidth]{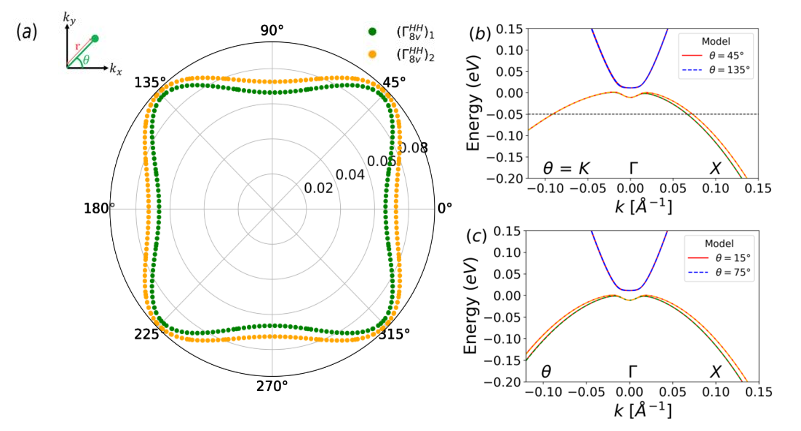}
    \caption{(a) The isoenergetic surface at $E = -0.05$ eV of the hybridized $\Gamma_{8v}^{HH}$ bands: $(\Gamma_{8v}^{HH})_{1}$ and $(\Gamma_{8v}^{HH})_{2}$ obtained using $H_{\mathrm{no\,BIA}}$ in radial coordinates i.e $(r,\,\theta)$ on the $k_{z}$ = 0 plane, where $\vec{k}\,=\,(r\cos\theta,\,r\sin\theta)$. Here $\theta$ represents the angle subtended by a vector $\vec{k}$ with the $k_{x}$ axis and $r$ represents the magnitude of $\vec{k}$. Comparison between the electronic band structure calculated along the $\theta$-$\Gamma$-$X$ path, for (b) $\theta$ = 45$\degree$ and 135$\degree$ (both equivalent to $\Gamma$-$K$ direction), where the dotted line represents the energy at which the isoenergetic surface in (a) has been constructed, and (c) $\theta$ = 15$\degree$ and $\theta$ = 75$\degree$.}
 \label{fig.3}
\end{figure*}

\par
Along the $\Gamma$-$K$ and $\Gamma$-$L$ directions, $k_{x} = k_{y} = \lambda$, where $\lambda = k / \sqrt{2}$ and $\lambda = k / \sqrt{3}$, respectively, we do not consider $k_{z}$, as $\epsilon_{xx} - \epsilon_{yy} = 0$ (Eqns.~\eqref{cancelZ88} and \eqref{cancelZ87}). Thus, we obtain the expressions
\begin{equation}
H_{C_{4}}^{8v,8v} = C_{4}(\epsilon_{yy} - \epsilon_{zz})\lambda(J_{x} - J_{y}),
\end{equation}
\begin{equation}
H_{C_{4}}^{8v,7v} = \frac{3}{2}C_{4}(\epsilon_{yy} - \epsilon_{zz})\lambda(U_{x} - U_{y}).
\end{equation}
We again consider the case of $k = 0.1$ $\mathrm{\mathring{A}^{-1}}$, but now along the $\Gamma$-$K$ path. In this case, $E_1 = -0.74$ $\mathrm{meV}$, $E_2 = 0.63$ $\mathrm{meV}$ and $\Delta E_{C_{4}} = 1.37$ $\mathrm{meV}$. The value of $\Delta E_{C_{4}}$ is rather small, making the energy bands indistinguishable. This implies that the $C_4$ strain terms induce negligible splitting between the $\Gamma_{8v}^{HH}$ bands along the $\Gamma$-$K$ and $\Gamma$-$L$ directions (see also Appendix section~\ref{app:high-symmetry-lines}).
\par
\begin{figure*}[tpb]
 \includegraphics[width=0.9\textwidth]{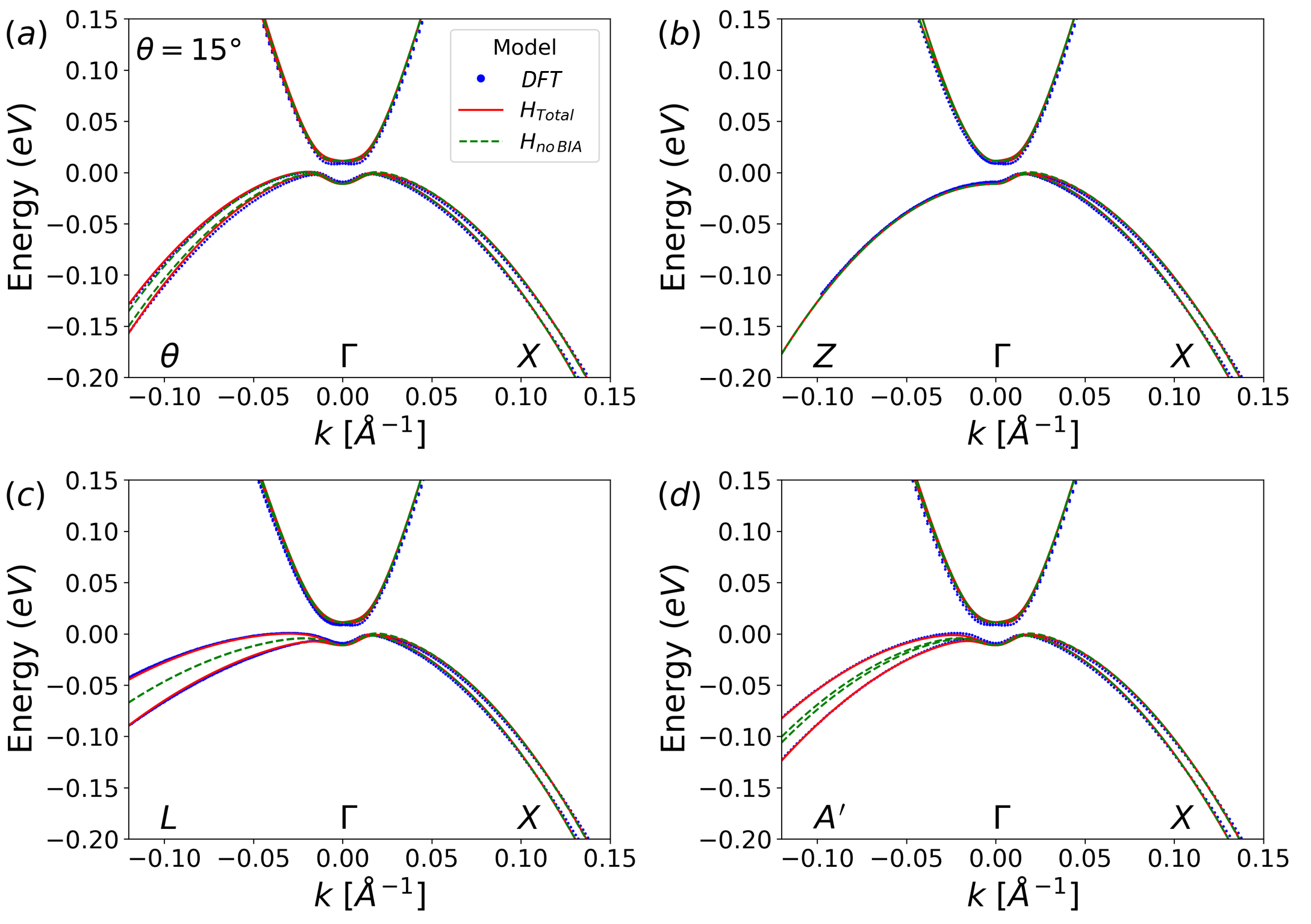}
    \caption{Fit of the 8x8 model Hamiltonian to the electronic band structure calculated using DFT with BIA ($H_{\mathrm{Total}}$) and without BIA ($H_{\mathrm{no\,BIA}}$) along the (a) $\theta$-$\Gamma$-$X$ path at $\theta$ = 15$\degree$ (b) $Z$-$\Gamma$-$X$ (c) $L$-$\Gamma$-$X$ and (d) an arbitrary path inclined at 57.65$\degree$ to the $k_z$ direction whose in-plane projection forms an angle of 31.64$\degree$ with the $k_x$ axes}
 \label{fig.4}
\end{figure*}

To confirm the above hypothesis we fit $H_{\mathrm{no\,BIA}}$ to our DFT results along the $K$-$\Gamma$-$X$ path (Fig. \ref{fig.2}(b)). We find that along the $\Gamma$-$X$ direction, the splitting is primarily induced by the $C_4$ strain terms, whereas in the $\Gamma$-$K$ direction, negligible splitting is induced by the $C_4$ strain terms. The same result is obtained when we fit $H_{\mathrm{no\,BIA}}$ to our DFT band structure calculated along the $L$-$\Gamma$-$X$ path (Fig. \ref{fig.4}(c)).
\par
Now, to study the effects of BIA on the energy eigenvalues we ignore $H_{C_4}$, thus our model Hamiltonian is $H_{\mathrm{no}\,C_{4}} = H_{\mathrm{Kane}} + H_{\mathrm{Pikus\text{--}Bir}} + H_{\mathrm{BIA}}$. To gauge the effects of $H_{\mathrm{BIA}}$ on the band splitting, we again apply the degenerate perturbation theory using $\ket{\Psi_{n = 1,2}}$. Like the case of $C_4$ strain terms, we again define a parameter $\Delta E_{\mathrm{BIA}}$ which describes the splitting induced by BIA and is equivalent to the difference between the eigenvalues of $H_{\mathrm{BIA}}$ represented in terms of the $\ket{\Psi_{n = 1,2}}$ basis. 
\par
For $k = 0.1$ $\mathrm{\mathring{A}^{-1}}$ on the $\Gamma$-$X$ path, we find that the eigenvalues are $E_1 = -0.77$ $\mathrm{meV}$, $E_2 = -0.17$ $\mathrm{meV}$, and $\Delta E_{\mathrm{BIA}} = 0.60$ $\mathrm{meV}$. Since this value is negligible, the $\Gamma_{8v}^{HH}$ bands are indistinguishable. We can thus confirm that the splitting produced along this path is primarily due to the $C_4$ strain terms, owing to the much larger value of $\Delta E_{C_{4}} = 14.46$ $\mathrm{meV}$. 

From our analysis of the $C_{4}$ strain terms above, it becomes evident that the BIA of the \ce{HgTe} lattice is responsible for band splitting along paths where the splitting induced by the $C_4$ terms is negligible. This hypothesis is strongly supported by our calculated eigenvalues at $k = 0.1$ $\mathrm{\mathring{A}^{-1}}$ along the $\Gamma$-$K$ direction: $E_1 = -27.9$ $\mathrm{meV}$, $E_2 = 9.7$ $\mathrm{meV}$ and $\Delta E_{\mathrm{BIA}} = 37.6$ $\mathrm{meV}$. Owing to the large value of $\Delta E_{\mathrm{BIA}}$ as compared to $\Delta E_{C_{4}} = 1.37$ $\mathrm{meV}$, we confirm that the BIA terms are primarily responsible for band splitting. To confirm our hypothesis, we fit $H_{\mathrm{no}\,C_{4}}$ to our DFT data along the $K$-$\Gamma$-$X$ path (Fig. \ref{fig.2}(c)) and find that the splitting induced by BIA terms along the $\Gamma$-$K$ direction is indeed much greater than the $C_4$ strain-induced splitting, thus making it the dominant cause of band splitting. The same can also be stated for the $\Gamma$-$L$ direction (Fig. \ref{fig.4}(c)). This fit also confirms that the splitting along the $\Gamma$-$X$ direction results primarily due to $H_{C_4}$, as $H_{\mathrm{no}\,C_{4}}$ produces negligible band splitting, rendering the $\Gamma_{8v}^{HH}$ bands indistinguishable.

\subsection{Band splitting in the $k_{z} = 0$ plane}\label{bandsplitting}

Prior to this study, the $C_4$ strain terms were not included in \kdotp{} models that discussed the band structure of strained \ce{HgTe} \cite{ruan2016symmetry,PhysRevB.72.035321,Leubner2017,mahler2021massive,zhang2001rashba,becker2000band,zhang2004effective,chen2019strain}. This can be attributed to the lack of experimental evidence of strain-induced sub-band splitting in \ce{HgTe} prior to \cite{PhysRevB.107.L121102} and the lower magnitude of band splitting induced by strain.
\par
To better understand the behaviour of the band splitting induced by $C_{4}$ strain term (in the absence of BIA), we plot the isoenergetic surface at $E = -0.05$ eV of the hybridized $\Gamma_{8v}^{HH}$ bands: $(\Gamma_{8v}^{HH})_{1}$ and $(\Gamma_{8v}^{HH})_{2}$ obtained using $H_{\mathrm{no\,BIA}}$ in radial coordinates i.e ($\theta$,\,$r$) on the $k_{z}$ = 0 plane. The resultant band dispersions obtained from $H_{\mathrm{no\,BIA}}$ have been depicted in Fig. \ref{fig.3}(a). This dispersion reveals that the magnitude of band splitting is maximum along the $\theta = 0\degree$ i.e the $k_{x}$ axis ($\Gamma$-$X$). As we move towards $\theta = 45\degree$, i.e., the $\Gamma$-$K$ direction, the splitting between the $\Gamma_{8v}^{HH}$ bands decreases and is minimum at $\theta = 45\degree$. As we move away from $\theta = 45\degree$, the splitting again increases and again becomes maximum at $\theta = 90\degree$. This implies that the band splitting induced by the $C_{4}$ strain term in the $k_{z} = 0$ plane is consistent with the four-fold rotational symmetry of the \ce{HgTe} lattice in the same plane. We verify this observation by comparing the electronic band structure at $\theta = 45\degree$ and $\theta = 225\degree$, which we find to be identical (Fig. \ref{fig.3}(b)). The resultant dispersion is also symmetric about the $\theta = 45\degree$ axis i.e the $\Gamma$-$K$ direction. This can be verified by comparing the electronic band structure at $\theta = 15\degree$ and $\theta = 75\degree$, which we also find to be identical (Fig. \ref{fig.3}(c)).  
%\vspace{-0.5 cm}

\subsection{Competition between $C_{4}$ strain and BIA terms to induce band splitting}

%\vspace{-0.3 cm}
The $C_4$ strain term has been used previously to describe the offset in quantum resonance energies obtained from cyclotron spectra in strained III-V semiconductors such as InSb \cite{ranvaud1979quantum,silver1992strain}. However, these studies have not established how the $C_4$ term would influence the band splitting in the presence of BIA as well as the varying magnitude of $C_4$-induced sub-band splitting along various crystallographic directions. Furthermore, these studies utilized band structures calculated by the primitive empirical pseudopotential method with no experimental input. Here, we probe the interplay between the $C_4$ strain and BIA terms by fitting our model to the band structure calculated along different $k$-paths using DFT to understand when each of these terms dominates the band splitting mechanism.

\begin{figure*}[t!]
 \includegraphics[width=1.0\textwidth]{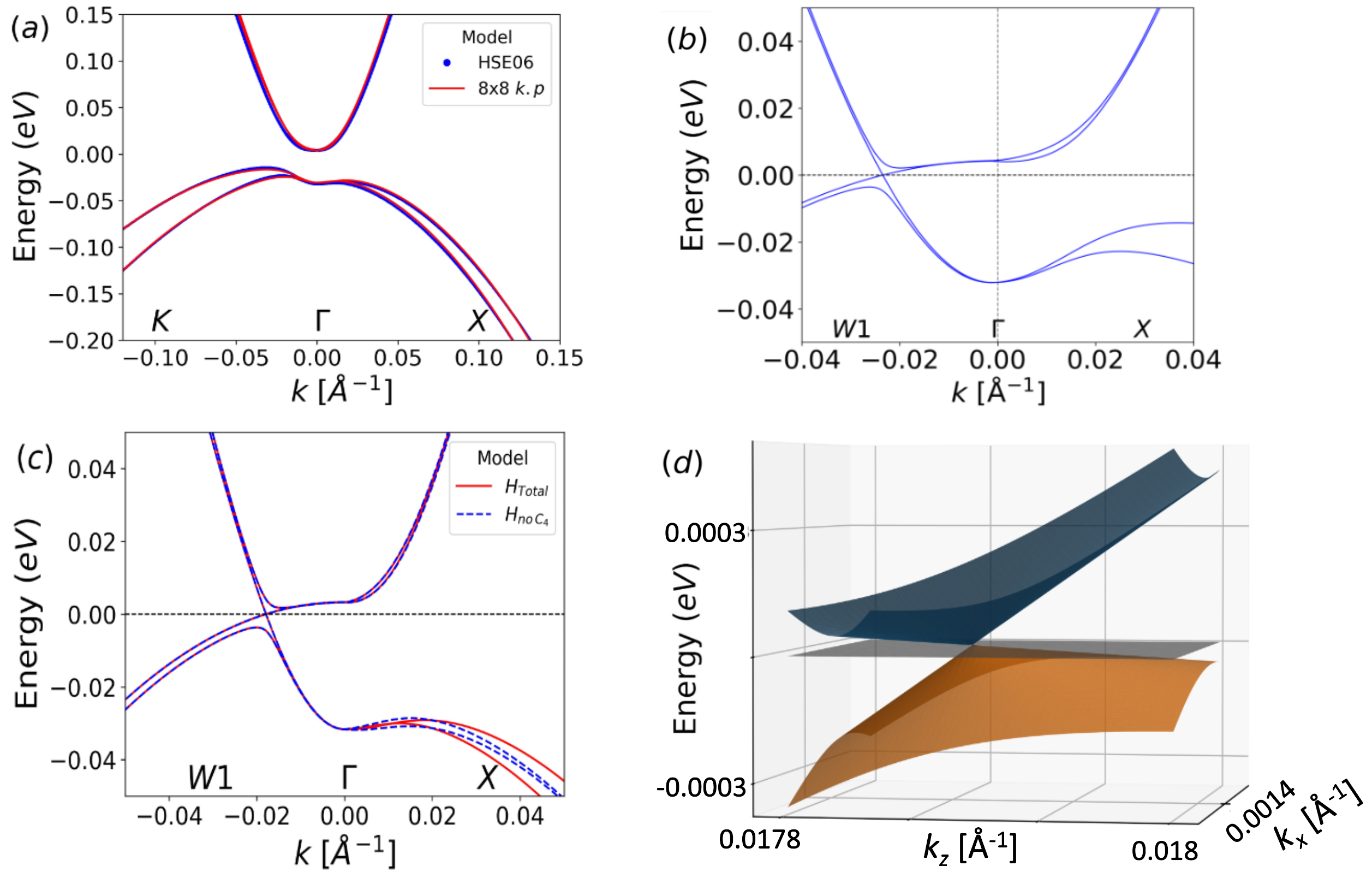}
    \caption{(a) A fit of $H_{\mathrm{Total}}$ to the DFT electronic structure in the Weyl semimetal state along the $K$-$\Gamma$-$X$ path.(b) The DFT band dispersion obtained along the $K$-$\Gamma$-$W1$ path, where $\Gamma$-$W1$ represents the line through the origin and one of the Weyl points. (c) A comparison of the \kdotp{} band structure obtained using $H_{\mathrm{Total}}$ and $H_{\mathrm{no}\,C_{4}}$ along the $W1$-$\Gamma$-$X$ path (d) The 3D band structure in the $k_{y} = 0$ plane depicting a tilted type-1 Weyl cone and at the Fermi surface.}
 \label{fig.5}
\end{figure*}

To gauge the competition between the $C_4$ strain and BIA terms in inducing band splitting of the $\Gamma_{8v}$ bands, we fit $H_{\mathrm{no\,BIA}}$ (includes $C_4$ terms but no BIA) and $H_{\mathrm{Total}}$ (includes both $C_4$ and BIA terms) to our DFT electronic band structure (Fig. \ref{fig.4}). We summarize the competition between the $C_4$ and BIA terms by considering four cases, namely: an arbitrary path inclined at 15$\degree$ to the $k_x$ axes, a path along the $k_z$ axes, the high symmetry $\Gamma$-$L$ direction and an arbitrary path $A'$ inclined at 57.65$\degree$ to the $k_z$ direction whose in plane projection forms an angle of 31.64$\degree$ with the $k_x$ axis. In the first case (Fig. \ref{fig.4}(a)) we find that a significant amount of the band splitting can be attributed to the $C_4$ strain terms as compared to the BIA terms. The dominance of $C_4$ terms in the band splitting phenomenon arises due to the proximity of this path to the $k_x$ axes where band splitting arises primarily due to the $C_4$ strain terms. Along the $k_z$ direction (Fig. \ref{fig.4}(b)) there is no band splitting. Though one would normally attribute splitting along this direction to the $C_4$ terms, the $C_4$ terms affiliated with the $k_z$ direction cancel out due to the isotropic nature of strain in the $k_z = 0$ plane, resulting in no splitting. The case of the $\Gamma$-$L$ high symmetry path (Fig. \ref{fig.4}(c)) is identical to the $\Gamma$-$K$ path as the band splitting arises completely due to the BIA terms, the cause of which has been discussed in Section \ref{origin}. On considering an arbitrary path $\Gamma$-$A'$ (Fig. \ref{fig.4}(d)) we find that the $C_4$ strain terms have a negligible contribution to the band splitting whereas the BIA terms have a substantial contribution to the band splitting. It can be estimated that the proximity of $\Gamma$-$A'$ to the $\Gamma$-$L$ high symmetry path is what causes the BIA terms to dominate the band splitting phenomenon.
%\vspace{-0.25 cm}

\subsection{The Weyl semimetal state}

\begin{figure*}[t!]
 \includegraphics[width=1.05\textwidth]{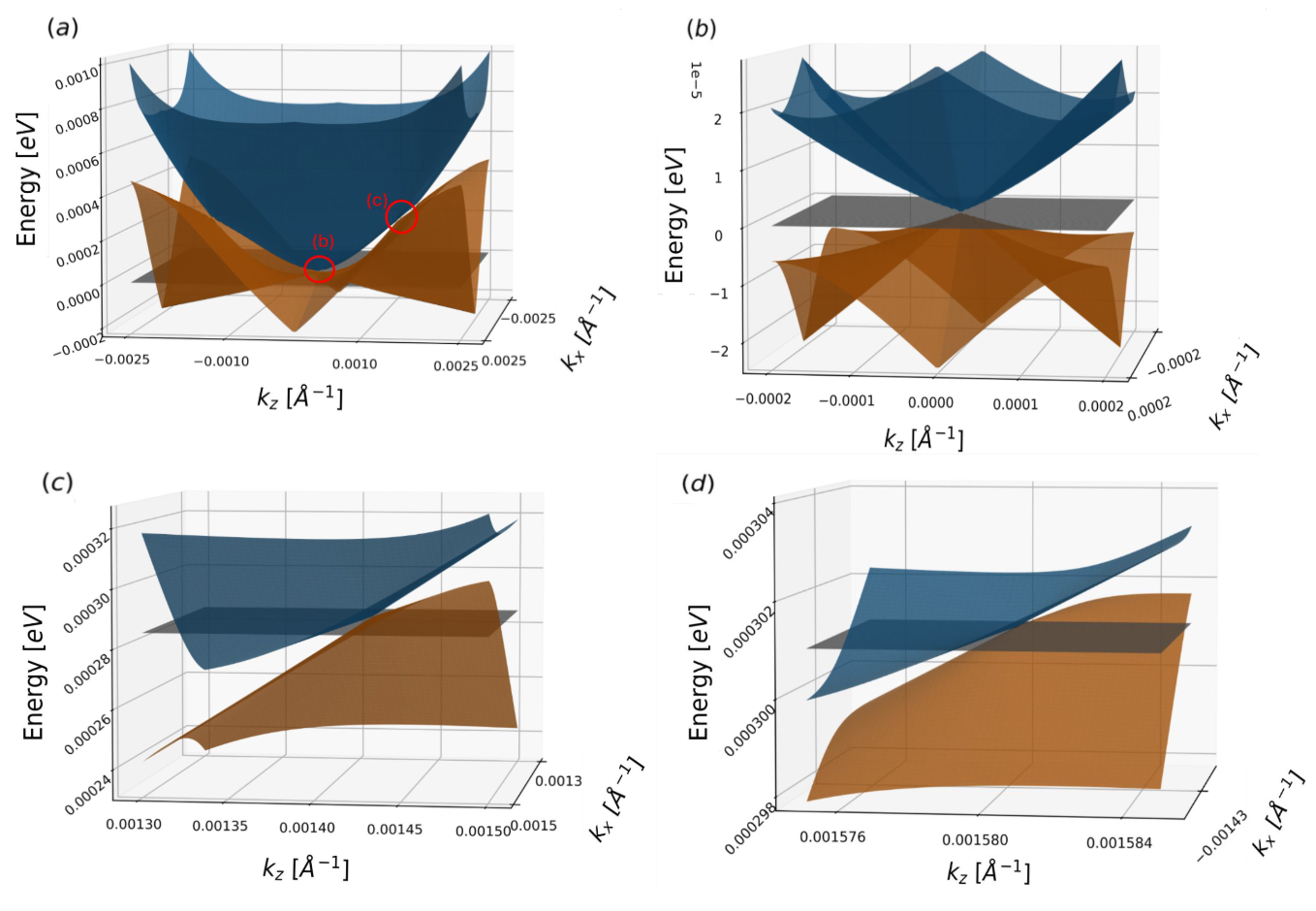}
    \caption{The 3D band structure in the $k_{y} = 0$ plane depicting (a) the different types of band crossings at 0\% strain labelled (circled in red) as (b) which shows a topologically trivial band crossing at the $\Gamma$ point and (c) which shows the type-2 Weyl dispersion. (d) We also find a type-2 Weyl dispersion at a surface isoenergetic to the Weyl points at -0.001\% (and 0.001\%). Surfaces isoenergetic to the Weyl points have been depicted as gray planes.}
 \label{fig.6}
\end{figure*}
\begin{table}[t]
\centering
\begin{tabular}{|l|l|l|l|}
\hline
Model & $k_x$ ($\mathrm{{\mathring{A}}}^{-1}$) & $k_y$ ($\mathrm{{\mathring{A}}}^{-1}$) & $k_z$ ($\mathrm{{\mathring{A}}}^{-1}$)\\ \hline
DFT-HSE06 & 0.001111 & -0.000511 & 0.018479\\ \hline
$H_{\mathrm{Total}}$ & 0.001438 & 0.000000 & 0.017871\\ \hline
$H_{\mathrm{no\,C_{4}}}$ & 0.001433 & 0.000000 & 0.017844\\ \hline
\end{tabular}
\caption{\label{table2} A comparison of the position of a Weyl point obtained at $\varepsilon_{xx} = -0.5\%$ using DFT-HSE06 and using the \kdotp{} Hamiltonians $H_{\mathrm{Total}}$ and $H_{\mathrm{no\,C_{4}}}$.}
\end{table}
Prior work \cite{ruan2016symmetry,chen2019strain,PhysRevB.87.045202,PhysRevLett.132.016603,PhysRevResearch.4.023114,kharitonov2022evolution,PhysRevX.9.031034,AravindnathEA2025} has demonstrated that \ce{HgTe} can be coaxed to attain a Weyl semimetal state when compressed, where the Weyl cones are located in the $k_x = 0$ and $k_y = 0$ plane. Here, we demonstrate that including the $C_{4}$ strain terms in our model Hamiltonian allows us to obtain a robust Weyl semimetal state, with Weyl point locations in the $k_y = 0$ plane at the Fermi level that are consistent with those obtained in the absence of these terms (see Table \ref{table2}). At extremely low strain, the system realizes a type-II Weyl semimetal phase, in which symmetry constrains the Weyl nodes to lie in the $k_x = 0$ and $k_y = 0$ planes. As the strain becomes more compressive, the phase diagram evolves and a transition to a tilted type-I Weyl semimetal phase occurs. In this regime, the symmetry constraints are relaxed and the Weyl nodes are no longer restricted to $k_x = 0$, allowing for finite $k_x = 0$ coordinates. Quantification of the tilt of these Weyl cones to confirm that they are indeed tilted type-I has been carried out in section \ref{quant} of the Appendix. The values reported in Table 3 correspond to $\varepsilon_{xx} = - 0.5\%$, i.e., within the tilted type-I region of the phase diagram. The small deviations from $k_y = 0$ observed in the DFT results originate from numerical and methodological aspects discussed below. To determine the position of Weyl points using our model Hamiltonian $H_{\mathrm{Total}}$, we first fit our model Hamiltonian to our DFT electronic band structure calculated along the $K$-$\Gamma$-$X$ path. We then use our fit to the DFT data (Fig. \ref{fig.5}(a)) to find the location of Weyl points.
A minimal change in the position of Weyl points is observed for $H_{\mathrm{Total}}$ and $H_{\mathrm{no}\,C_{4}}$ (see Table~\ref{table2}), which implies that the inclusion of the $C_4$ strain terms have no effect on the Weyl semimetal state. Therefore, models that ignore the $C_4$ strain terms are still suitable to probe the Weyl semimetal phase. However, on comparing the Weyl point location obtained by our model Hamiltonian to that obtained by Wannier interpolation of the DFT band structure, we find that the Weyl point location predicted by our functional does not lie in the $k_y = 0$ plane. This can be attributed to the loss of symmetries during the Wannier interpolation of the DFT band structure. 
\par
This demonstrates that the role of the $C_4$ strain terms must be distinguished between two physically distinct regimes. In the TI regime, where the band structure is characterized by sub-band splitting and camel-back features, the $C_4$ terms play a crucial role. As demonstrated in Fig. \ref{fig.2}, the splitting along certain directions is dominated by the $C_4$ contribution, particularly in cases where the bulk inversion asymmetry (BIA) terms are weak.
\par
However, in the Weyl semimetal regime, the situation is qualitatively different. Once the band inversion and symmetry breaking (arising primarily from strain and BIA) drive the formation of Weyl nodes, the existence of the nodes themselves does not rely on the presence of $C_4$ terms. Instead, the $C_4$ contributions act as higher-order corrections that modify the dispersion and anisotropy of the Weyl cones without changing their topological character.
\par
In this sense, the $C_4$ terms are crucial for accurately capturing the band structure in the vicinity of the $\Gamma$ point and for reproducing sub-band splitting, but they are not required for the existence of the Weyl phase itself. The Weyl nodes remain robust against the removal of $C_4$ terms, although their precise locations and anisotropies may be quantitatively altered.
\par
Next, we study the change in the \ce{HgTe} band structure across the compressive strain regime using $H_{\mathrm{Total}}$ by extrapolating our Weyl semimetal fit parameters obtained at $\epsilon_{xx}$ = $-0.5$\% to other strain values. This results in three distinct strain regions that host a Weyl semimetal state. To classify the Weyl semimetal state we use one of the methods described in \cite{li2017evidence}, wherein we plot the 3D band structure in the $k_{y} = 0$ plane and probe Weyl band crossings at the Fermi surface for charge pockets. The type-2 Weyl semimetal state can be characterized by a significant tilt of the Weyl cones such that they cut the Fermi surface (or isoenergetic surface of the Weyl point) to form charge pockets. If there is a considerable tilt of the Weyl cones, but no charge pockets at the Fermi surface, the phase is classified as a tilted type-1 Weyl semimetal. The complete absence of a tilt of Weyl cones implies that the phase is an ideal type-1 Weyl semimetal.
\par
%\vspace{-0.001 cm}

\bigskip\noindent\emph{The high strain region}---
At large compressive strains such as $-0.5$\% we observe a tilted type-1 Weyl semimetal state \cite{zhang2018berry} with and without $C_4$ terms (Fig. \ref{fig.5}(c)). This is in stark contrast to prior work \cite{ruan2016symmetry,chen2019strain} that predicts an ideal type-1 Weyl semimetal (no tilt). Fermi level analysis of the 3D band structure (Fig. \ref{fig.5}(d)) confirms the presence of a tilted type-1 Weyl semimetal due to the absence of charge pockets and considerable inclination of the Weyl cone. Literature \cite{zhang2018berry} that has studied the Weyl semimetal state in various materials has demonstrated an increase in the Berry curvature dipole of a system with the tilting of Weyl cones. Since strained \ce{HgTe} can display a Berry curvature dipole \cite{chen2019strain}, realizing this phase will prove useful for enhancing it for use in future applications. Furthermore, the ability of our model to predict tilted type-1 Weyl cones, makes it useful for studying the superconducting diode effect observed in such tilted Weyl semimetals \cite{PhysRevB.109.064511}.
\par
Transition from type-1 to type-2: On lowering the magnitude of compressive strain applied on the $\mathrm{HgTe}$ lattice, the Weyl nodes shift above the Fermi level and experience an increase in their tilt. This is consistent with prior work\cite{ruan2016symmetry}, wherein the Weyl nodes shift slightly from the Fermi level and the $k_x = 0$ and $k_y = 0$ planes.
\par
%\vspace{-0.001 cm}
\bigskip\noindent\emph{The unstrained (0\%) state}---
In the unstrained state we observe two types of crossings (Fig. \ref{fig.6}(a)). The first is centered at the $\Gamma$ high symmetry $k$-point on the Fermi surface (Fig. \ref{fig.6}(b)), whereas the second kind are significantly tilted and lie above the Fermi level (Fig. \ref{fig.6}(c)). Previous work \cite{ruan2016symmetry,chen2019strain,AravindnathEA2025} has demonstrated the existence of nodal lines that contain Weyl points in the $k_{x} = 0$ and $k_{y} = 0$ planes at 0\% strain which converge at $\Gamma$. This implies that the first crossing is topologically trivial and not a Weyl point. To accurately classify the nature of the second type of band crossing, we construct a surface that is isoenergetic with the observable Weyl points. The presence of charge pockets at the isoenergetic surface implies that unstrained \ce{HgTe} also hosts a type-2 Weyl semimetal state.
\par
%\vspace{-0.07028 cm}
\bigskip\noindent\emph{The low strain region}---
At very small strain, such as $-0.001$\% (and $0.001$\%), we observe a substantial increase in the inclination of the Weyl cones in Fig.~\ref{fig.6}(d) (see Appendix section \ref{stopo} ), as well as the energy of the Weyl points, which now lie above the Fermi level. The significant tilt of the Weyl cones results in charge pockets being observed at their isoenergetic surface implying that the Weyl semimetal state is of type-2. 
%\vspace{-0.5 cm}
\par
Transition from type-2 to TI state: On increasing the tensile strain beyond a threshold value, the type-2 Weyl nodes of opposite parities merge and annihilate each other to form a robust TI state as described in prior work\cite{ruan2016symmetry,mahler2021massive,Leubner2017}.

\section{Conclusions}
\label{conc}
%\vspace{-0.3 cm}
We have studied the effects of strain on the sub-band splitting mechanism in the 3D topological insulator \ce{HgTe} by fitting our \kdotp{} model to a state of the art DFT calculations able to quantitatively describe the photoemission spectra. The inclusion of the higher order $C_4$ strain terms in our model, which compete with the intrinsic BIA of the \ce{HgTe} lattice, is crucial for understanding the sub-band degeneracy breaking and results in a $k$-path specific sub-band splitting phenomenon depending on whether strain or BIA dominates. 
\par
We apply our extended \kdotp{} model to study the topological phase transition in the Weyl semimetal state. Our model is consistent with the phase diagram predicted in prior research with the exception of the high compressive strain region, where we observe a tilted type-1 Weyl semimetal instead of an ideal type-1 Weyl semimetal. Such a tilted Weyl semimetal state would be suitable for applications that require a Berry curvature dipole which results from this tilt. 
\par
Our work provides a more accurate insight as to how the incorporation of symmetry breaking terms such as BIA and $C_4$ into the \kdotp{} model Hamiltonian influence the camel's back formation in the tensile strain phase and the topological phase transition towards a Weyl phase as a function of strain.

\section*{Acknowledgements}
We thank Domenico Di Sante for insightful discussions.
E.K. acknowledges Laurens W.\linebreak[4] Molenkamp and Hartmut Buhmann for the hospitality provided at the Physikalisches Institut (EP3), Universit\"{a}t W\"{u}rzburg, and at the Institute for Topological Insulators. 

% TODO: include funding information
\paragraph{Funding information}
E.K. acknowledges financial support from the German Academic Exchange Service (DAAD) through the WISE program.
G.M., P.M.F. and G.P. acknowledge CINECA under the ISCRA initiative. G.M. acknowledges funding by the European Union (ERC, DELIGHT, 101052708). Views and opinions expressed are however those of the author(s) only and do not necessarily reflect those of the European Union or the European Research Council. Neither the European Union nor the granting authority can be held responsible for them.
P.M.F. and G.P. thanks the Director and the Computing Network Service of the Laboratori Nazionali del Gran Sasso (LNGS-INFN). This research used resources of the LNGS HPC cluster realised in the framework of Spoke 0 and Spoke 5 of the ICSC project - Centro Nazionale di Ricerca in High Performance Computing, Big Data and Quantum Computing, funded by the NextGenerationEU European initiative through the Italian Ministry of University and Research, PNRR Mission 4, Component 2: Investment 1.4, Project code CN00000013 - CUP I53C21000340006.
G.P. acknowledges fundings from the European Union - NextGenerationEU under the Italian Ministry of University and Research (MUR) National Innovation Ecosystem grant ECS00000041 - VITALITY - CUP E13C22001060006.
G.S. and W.B. acknowledge financial support from the Deutsche Forschungsgemeinschaft
(DFG, German Research Foundation) in the project SFB 1170 \emph{ToCoTronics}
(Project ID 258499086) and in the W\"urzburg-Dresden Cluster of Excellence on Complexity and Topology in Quantum Matter \emph{ct.qmat} (EXC 2147, Project ID 390858490).

\begin{appendix}
\numberwithin{equation}{section}
\setcounter{figure}{0}
\renewcommand{\figurename}{Fig.}
\renewcommand{\thefigure}{S\arabic{figure}}
\section{Definition of the matrix Hamiltonian}
\subsection{The Kane Hamiltonian $H_{\mathrm{Kane}}$}\label{hkanedef}
$H_{\mathrm{Kane}}$ has been constructed in terms of the 8 orbital basis mentioned in Section \ref{k.p} and can be described as
\begin{equation}
H_{\mathrm{Kane}} = 
\begin{bmatrix}
H^{6c,6c} & H^{6c,8v} & H^{6c,7v}\\
(H^{6c,8v})^{\dag} & H^{8v,8v} & H^{8v,7v}\\
(H^{6c,7v})^{\dag} & (H^{8v,7v})^{\dag} & H^{7v,7v}\\
\end{bmatrix}.
\end{equation}
Each individual block can be expanded as follows,
\begin{equation}
H^{6c,6c} = 
\begin{bmatrix}
E_{v} + E_{0} + \frac{{{\hbar}^2}{k}^{2}}{2m'} & 0\\
0 & E_{v} + E_{0} + \frac{{{\hbar}^2}{k}^{2}}{2m'}\\
\end{bmatrix},
\end{equation}
\begin{equation}
H^{6c,7v} = 
\begin{bmatrix}
-\frac{1}{\sqrt{3}}Pk_{z} & -\frac{1}{\sqrt{3}}Pk_{-}\\
-\frac{1}{\sqrt{3}}Pk_{+} &  \frac{1}{\sqrt{3}}Pk_{z}\\
\end{bmatrix},
\end{equation}
\begin{equation}
H^{6c,8v} = 
\begin{bmatrix}
-\frac{1}{\sqrt{2}}Pk_{+} & \sqrt{\frac{2}{3}}Pk_{z} & \frac{1}{\sqrt{6}}Pk_{-} & 0\\
0 & -\frac{1}{\sqrt{6}}Pk_{+} & \sqrt{\frac{2}{3}}Pk_{z} & \frac{1}{\sqrt{2}}Pk_{-} \\
\end{bmatrix},
\end{equation}
\\
\begin{equation}
H^{8v,8v} = 
\begin{bmatrix}
U + V & S^{\dag} & R^{\dag} & 0\\
S & U - V & 0 & R^{\dag}\\
R & 0 & U - V & - S^{\dag}\\
0 & R & - S &  U + V\\
\end{bmatrix},
\end{equation}
% \\
% \hspace*{0cm}
%\vspace{0.5cm}
\begin{equation}
H^{8v,7v} = 
\begin{bmatrix}
-\frac{1}{\sqrt{2}}S^{\dag} & -\sqrt{2}R^{\dag}\\
-\sqrt{2}V & \sqrt{\frac{3}{2}}S^{\dag}\\
\sqrt{\frac{3}{2}}S & \sqrt{2}V\\
\sqrt{2}R & -\frac{1}{\sqrt{2}}S
\end{bmatrix},
\end{equation}
\begin{equation}
H^{7v,7v} = 
\begin{bmatrix}
E_{v} - \Delta_{0} - \frac{{{\hbar}^2}}{2m_{0}}\gamma_{1}{k}^{2} & 0\\
0 & E_{v} - \Delta_{0} - \frac{{{\hbar}^2}}{2m_{0}}\gamma_{1}{k}^{2}\\
\end{bmatrix},
\end{equation}
where 
\begin{equation}
U = E_{v} - \frac{\hbar^{2}\gamma_{1}}{2m_{0}}(k_{\parallel}^{2} + k_{z}^{2}),
\end{equation}
\begin{equation}
V = -\frac{\hbar^{2}\gamma_{2}}{2m_{0}}(k_{\parallel}^{2} - 2k_{z}^{2}),
\end{equation}
\begin{equation}
S = 2\sqrt{3}\frac{{\hbar}^2}{2m_{0}}\gamma_{3}k_{+}k_{z},
\end{equation}
\begin{equation}
R = \sqrt{3}\frac{{\hbar}^2}{2m_{0}}\gamma_{2}\hat{K} + 2i\sqrt{3}\frac{{\hbar}^2}{2m_{0}}\gamma_{3}k_{x}k_{y},
\end{equation}
and where $k^2 = k_{x}^2 + k_{y}^2 + k_{z}^2$, $k_{\parallel}^{2} = k_{x}^2 + k_{y}^2$, $k_{\pm} = k_{x} \pm ik_{y}$, and $\hat{K} = k_{x}^{2} - k_{y}^{2}$.

$E_{v}$ represents the valence band maxima, which amounts to about $0$ $\mathrm{eV}$ for the TI state and $-0.0136$ $\mathrm{eV}$ for the Weyl semimetal state. $E_{0}$ represents the energy gap between the $\Gamma_{6c}$ bands and the $\Gamma_{8v}^{LH}$ light hole (LH) bands (i.e, the bands corresponding to ${\ket{\Gamma_8,\frac{\pm{1}}{2}}}$). We find that $E_0 = -0.37$ $\mathrm{eV}$ for the TI state and $E_0 = -0.3913$ $\mathrm{eV}$ for the Weyl semimetal state. $\Delta_0$ represents the energy gap between the energy of the $\Gamma_{8v}$ HH bands at the $\Gamma$ point and the $\Gamma_{7v}$ bands. It amounts to about $0.87$ $\mathrm{eV}$ and $0.821$ $\mathrm{eV}$ for the topological insulator and Weyl semimetal state, respectively. $m'$ is related to the free electron mass $m_0$ via the expression
\begin{equation}
m' = \frac{m_0}{2F+1},
\end{equation}
where $F = 0.0$ \cite{PhysRevB.72.035321}.

$\gamma_{1}$, $\gamma_{2}$ and $\gamma_{3}$ represent the Luttinger parameters. For the TI state, $\gamma_{1}$ and $\gamma_{2}$ show negligible variation in their estimated mean values of about $3.804$ and $0.385$, respectively, whereas the mean value of $\gamma_{3} = 1.195$ varies by $\pm16\%$. Whereas, for the Weyl semimetal state, $\gamma_{1} = 4.1$, $\gamma_{2} = 0.57$ and $\gamma_{3} = 1.23$. Here $P$ represents the expectation value of the momentum operator $\hat{p}_a$ with the $s$ and $p_a$ orbitals where $a = (x, y, z)$ and can be written as $P = -\frac{\hbar}{m_0}\mel{s}{\hat{p}_a} {p_{a}}$.
For our calculations we set $P = 8.46$ $\mathrm{eV \mathring{A}}$.

\subsection{The strain Hamiltonian $H_{\mathrm{Pikus\text{--}Bir}}$}\label{hpikusbirdef}

In the absence of shear strain, $H_{\mathrm{Pikus\text{--}Bir}}$ can be represented in terms of the same basis as

\begin{equation}
\scriptstyle
H_{\mathrm{Pikus\text{--}Bir}} = 
{\arraycolsep=1pt
\begin{bmatrix}
T_{\epsilon} & 0 & 0 & 0 & 0 & 0 & 0 & 0\\
0 & T_{\epsilon} & 0 & 0 & 0 & 0 & 0 & 0\\
0 & 0 & U_{\epsilon} + V_{\epsilon} & 0 & R_{\epsilon} & 0 & 0 & -\sqrt{2}R_{\epsilon}\\
0 & 0 & 0 & U_{\epsilon} - V_{\epsilon} & 0 & R_{\epsilon} & \sqrt{2}V_{\epsilon} & 0\\
0 & 0 & {R_{\epsilon}} & 0 & U_{\epsilon} - V_{\epsilon} & 0 & 0 & -\sqrt{2}V_{\epsilon}\\
0 & 0 & 0 & {R_{\epsilon}} & 0 & U_{\epsilon} + V_{\epsilon} & \sqrt{2}R_{\epsilon} & 0\\
0 & 0 & 0 & \sqrt{2}V_{\epsilon} & 0 & \sqrt{2}R_{\epsilon} & U_{\epsilon} & 0\\
0 & 0 & -\sqrt{2}R_{\epsilon} & 0 & \sqrt{2}V_{\epsilon} & 0 & 0 & U_{\epsilon}\\
\end{bmatrix}},
\end{equation}
where 
\begin{equation}
T_{\epsilon} = c_{s}tr(\epsilon),
\end{equation}
\begin{equation}
U_{\epsilon} = a_{s}tr(\epsilon),
\end{equation}
\begin{equation}
V_{\epsilon} = \frac{1}{2}b_{s}(\epsilon_{xx} + \epsilon_{yy} - 2\epsilon_{zz}),
\end{equation}
\begin{equation}
R_{\epsilon} = \frac{\sqrt{3}}{2}b_{s}(\epsilon_{xx} - \epsilon_{yy})
\end{equation}
in which $tr(\epsilon) = \epsilon_{xx} + \epsilon_{yy} + \epsilon_{zz}$ (the trace of the strain tensor), and $R_{\epsilon} = 0$ as we consider the in-plane ($k_z = 0$) strain to be isotropic (i.e $\epsilon_{xx} = \epsilon_{yy}$). The coefficients $a_s$, $b_s$ and $c_s$ are treated as constants throughout our calculations and amount to $0.0$ $\mathrm{eV}$, $-1.5$ $\mathrm{eV}$ and $-3.83$ $\mathrm{eV}$, respectively.
\begin{figure*}[t!]
\centering
  \includegraphics[width=0.6\linewidth]{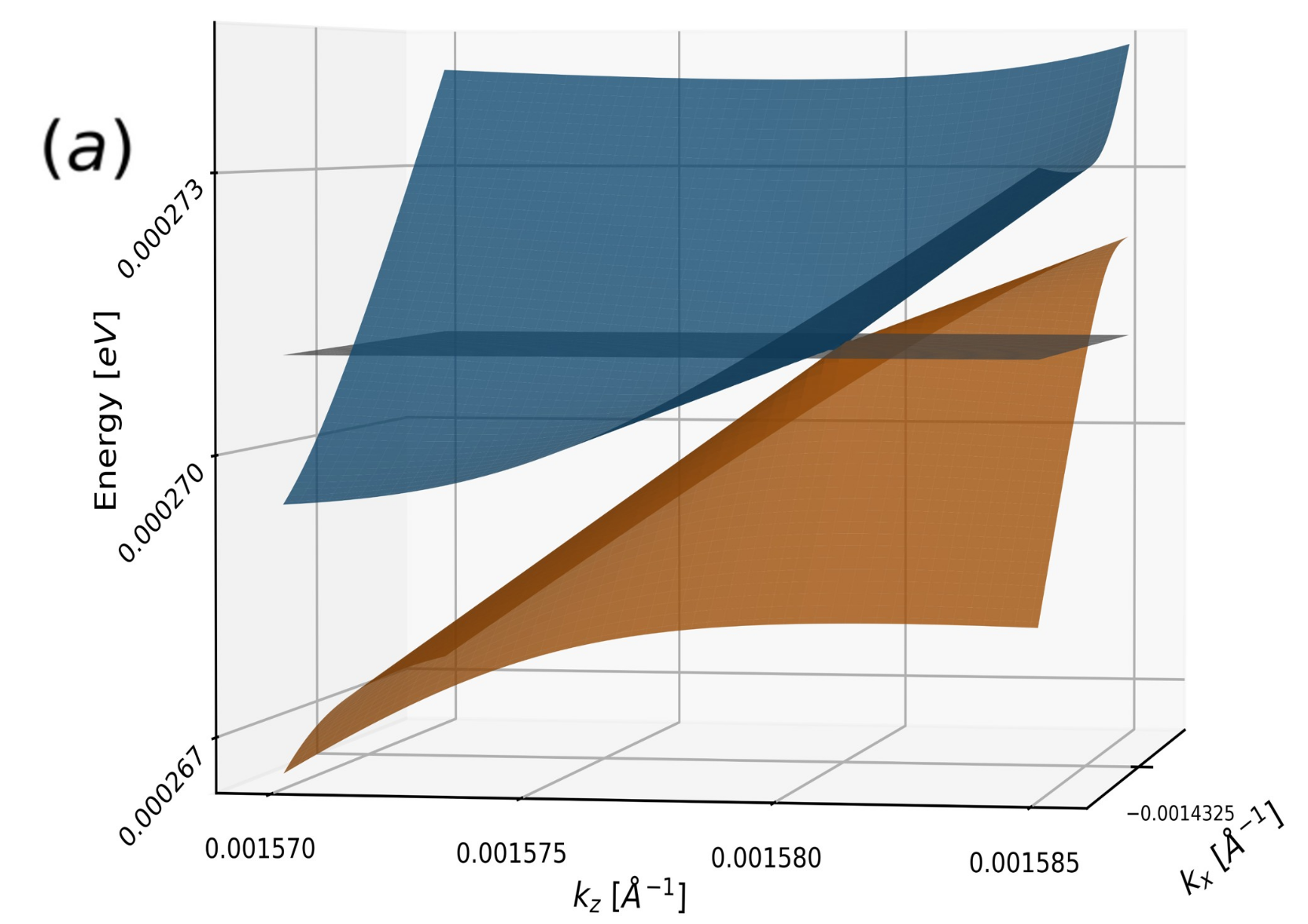}\\
\centering
  \includegraphics[width=0.6\linewidth]{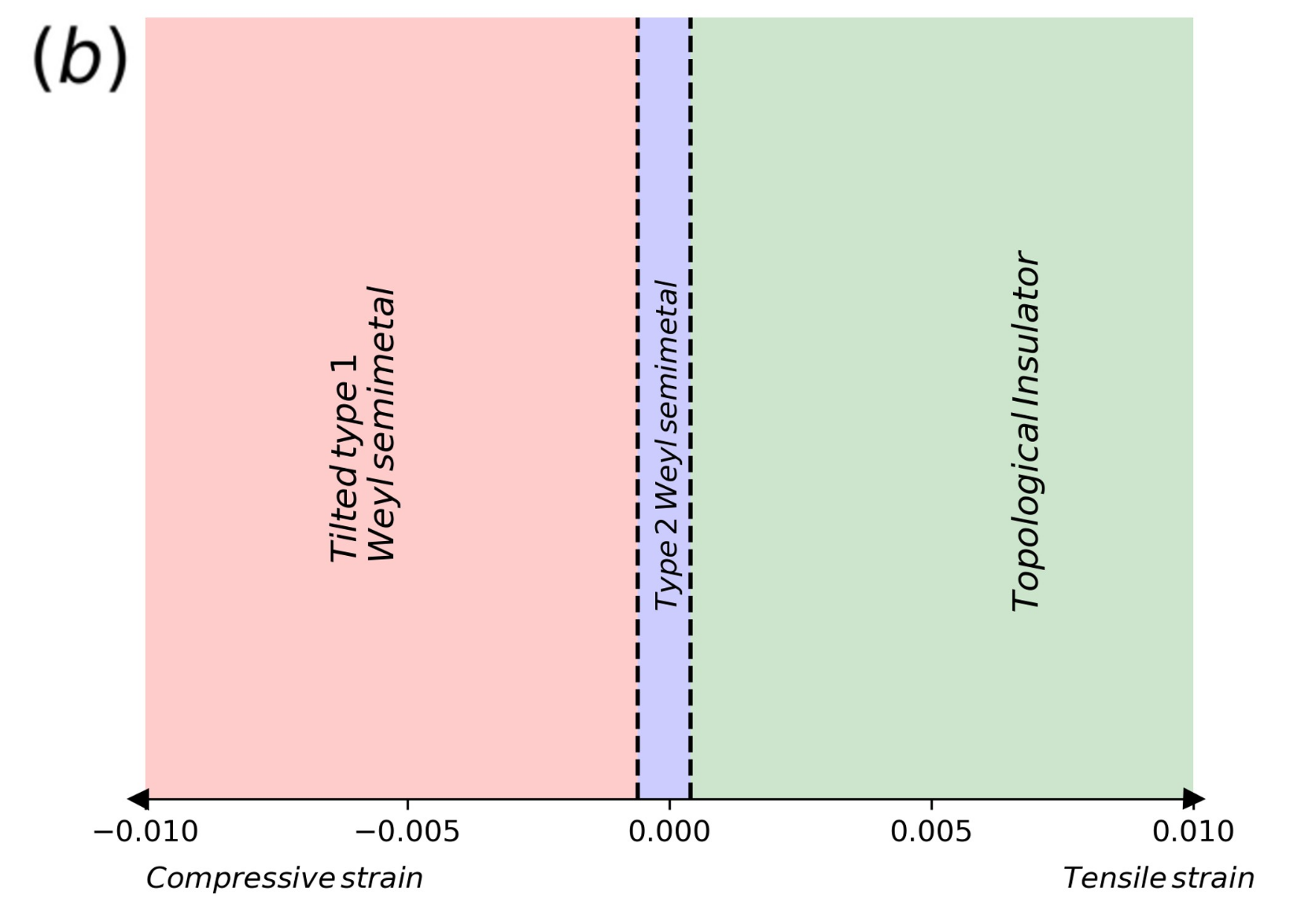}
\caption{(a) The 3D band structure in the $k_{y} = 0$ plane depicting the type-2 Weyl dispersion at a surface isoenergetic to the Weyl points (gray plane) at 0.001\%. (b) The topological phase diagram of \ce{HgTe} as a function of strain.}
 \label{fig.S1}
\end{figure*}

\subsection{The BIA Hamiltonian $H_{\mathrm{BIA}}$}\label{biadef}
$H_{\mathrm{BIA}}$ can be described as
\begin{equation}
H_{\mathrm{BIA}} = 
\begin{bmatrix}
  0 & H_{\mathrm{BIA}}^{6c,8v} & H_{\mathrm{BIA}}^{6c,7v}\\[1em]
  (H_{\mathrm{BIA}}^{6c,8v})^{\dag} & H_{\mathrm{BIA}}^{8v,8v} & H_{\mathrm{BIA}}^{8v,7v}\\[1em]
  (H_{\mathrm{BIA}}^{6c,7v})^{\dag} & (H_{\mathrm{BIA}}^{8v,7v})^{\dag} & 0\\[1em]
\end{bmatrix},
\end{equation}
where each block can be depicted in terms of the momentum ($k_x$, $k_y$, $k_z$) and coefficients $C$, $B_{8v}^{+}$, $B_{8v}^{-}$ and $B_{7v}$. Each of the above listed blocks can be described as

%\vspace{-0.2cm}
\begin{equation}
(H_{\mathrm{BIA}}^{6c,8v})^{\dag} = {\arraycolsep=1pt
\begin{bmatrix}
  \frac{1}{\sqrt{2}}B_{8v}^{+}k_{+}k_{z} & -\frac{1}{3\sqrt{2}}B_{8v}^{-}(k_{\parallel}^{2} - 2k_{z}^{2})\\
 +\frac{1}{\sqrt{6}}B_{8v}^{-}\hat{K} - i\sqrt{\frac{2}{3}}B_{8v}^{+}k_{x}k_{y} & \frac{1}{\sqrt{6}}B_{8v}^{+}k_{+}k_{z}\\
  \frac{1}{\sqrt{6}}B_{8v}^{+}k_{-}k_{z} & -\frac{1}{\sqrt{6}}B_{8v}^{-}\hat{K} - i\sqrt{\frac{2}{3}}B_{8v}^{+}k_{x}k_{y}\\
  \frac{1}{3\sqrt{2}}B_{8v}^{-}(k_{\parallel}^{2} - 2k_{z}^{2}) & +\frac{1}{\sqrt{2}}B_{8v}^{+}k_{-}k_{z}\\
\end{bmatrix}},
\end{equation}
%\vspace{-0.2cm}
\begin{equation}
H_{\mathrm{BIA}}^{6c,7v} = 
\begin{bmatrix}
-\frac{i}{\sqrt{3}}B_{7v}k_{x}k_{y} & -\frac{1}{\sqrt{3}}B_{7v}k_{+}k_{z}\\ 
+\frac{1}{\sqrt{3}}B_{7v}k_{-}k_{z} & \frac{i}{\sqrt{3}}B_{7v}k_{x}k_{y}\\
\end{bmatrix},
\end{equation}
\begin{equation}
H_{\mathrm{BIA}}^{8v,8v} = 
\begin{bmatrix}
0 & -\frac{1}{2}Ck_{+} & Ck_{z} & -\frac{\sqrt{3}}{2}Ck_{-}\\
-\frac{1}{2}Ck_{-} & 0 & \frac{\sqrt{3}}{2}Ck_{+} & - Ck_{z}\\
Ck_{z} & \frac{\sqrt{3}}{2}Ck_{-} & 0 & -\frac{1}{2}Ck_{+}\\
\frac{\sqrt{3}}{2}Ck_{+} & - Ck_{z} & -\frac{1}{2}Ck_{-} & 0 \\
\end{bmatrix},
\end{equation}
\begin{equation}
(H_{\mathrm{BIA}}^{8v,7v})^{\dag} = 
\begin{bmatrix}
\frac{1}{2\sqrt{2}}Ck_{-} & 0 & \frac{\sqrt{3}}{2{\sqrt{2}}}Ck_{+} & \frac{1}{\sqrt{2}}Ck_{z}\\
\frac{1}{\sqrt{2}}Ck_{z} & -\frac{\sqrt{3}}{2\sqrt{2}}Ck_{-} & 0 & -\frac{1}{2\sqrt{2}}Ck_{+}\\
\end{bmatrix},
\end{equation}

From our fit to the DFT data in the TI as well as the Weyl semimetal state, we find that $B_{8v}^{+}$, $B_{8v}^{-}$ and $B_{7v}$ remain constant with values of about $-10.646$ $\mathrm{eV {\mathring{A}}^{2}}$, $-1.377$ $\mathrm{eV {\mathring{A}}^{2}}$ and $10$ $\mathrm{eV {\mathring{A}}^{2}}$, respectively. We find that the mean value $C = 0.126$ $\mathrm{eV {\mathring{A}}}$, varies by about $\pm10\%$ along different paths in the Brillouin zone for the TI state and is about $C = 0.131$ $\mathrm{eV {\mathring{A}}}$ for the Weyl semimetal state.

\section{Definition of $J$ and $U$ matrices}\label{JUdef}
The angular momentum $J_{a}$ matrices, $a = (x, y, z)$ have been constructed in terms of the $\ket{j,m}$ basis corresponding to the $\Gamma_{8v}$ bands.
\begin{equation}
J_x = \frac{1}{2}
\begin{bmatrix}
0 & \sqrt{3} & 0 & 0\\
\sqrt{3} & 0 & 2 & 0\\
0 & 2 & 0 & \sqrt{3}\\
0 & 0 & \sqrt{3} & 0\\
\end{bmatrix},
\end{equation}
\begin{equation}
J_y = \frac{i}{2}
\begin{bmatrix}
0 & -\sqrt{3} & 0 & 0\\
\sqrt{3} & 0 & -2 & 0\\
0 & 2 & 0 & -\sqrt{3}\\
0 & 0 & \sqrt{3} & 0\\
\end{bmatrix},
\end{equation}
\begin{equation}
J_z = \frac{1}{2}
\begin{bmatrix}
3 & 0 & 0 & 0\\
0 & 1 & 0 & 0\\
0 & 0 & -1 & 0\\
0 & 0 & 0 & -3\\
\end{bmatrix},
\end{equation}
The $U_{a}$ matrices, $a = (x, y, z)$ are needed to describe the interactions between the $\Gamma_{8v}$ bands and the $\Gamma_{6c}$ or the $\Gamma_{7v}$ bands, respectively. They can be defined as 
\begin{equation}
U_x = \frac{1}{3\sqrt{2}}
\begin{bmatrix}
-\sqrt{3} & 0\\
0 & -1\\
1 & 0\\
0 & \sqrt{3}\\
\end{bmatrix},
\end{equation}
\begin{equation}
U_y = \frac{i}{3\sqrt{2}}
\begin{bmatrix}
\sqrt{3} & 0\\
0 & 1\\
1 & 0\\
0 & \sqrt{3}\\
\end{bmatrix},
\end{equation}
\begin{equation}
U_z = \frac{\sqrt{2}}{3}
\begin{bmatrix}
0 & 0\\
0 & 1\\
1 & 0\\
0 & 0\\
\end{bmatrix}.
\end{equation}

\section{Symmetry analysis of the Hamiltonian terms}
\label{app:symmetry}

\subsection{Invariant-level symmetry criterion and transformation rules}

A point-group operation $g$ is represented by an orthogonal matrix $R_g$ acting on real space.
The Hamiltonian is invariant under $g$ if
\begin{equation}
U_g\,H(\mathbf{k},\varepsilon)\,U_g^{-1}
=H(R_g\mathbf{k},\,R_g\,\varepsilon\,R_g^{T}),
\label{eq:symCriterion}
\end{equation}
where $U_g$ is the (double-group) representation on the band basis. In the invariant approach we do not need $U_g$ explicitly; it suffices to track how the building blocks transform.

The relevant objects transform as follows:
\begin{align}
\text{polar vector:} \quad & \mathbf{k}\mapsto R_g\mathbf{k}, \label{eq:kTransform}\\
\text{axial vector:} \quad & \mathbf{J}\mapsto (\det R_g)\,R_g\mathbf{J}, \label{eq:JTransform}\\
\text{rank-2 tensor:}\quad & \varepsilon \mapsto R_g\,\varepsilon\,R_g^{T}. \label{eq:epsTransform}
\end{align}
(Any operator triplet $\mathbf{U}=(U_x,U_y,U_z)$ transforming as an axial vector follows the same rule as $\mathbf{J}$.)

For inversion $\mathcal{I}$ one has $R_{\mathcal I}=-\mathbb{1}$ and $\det R_{\mathcal I}=-1$. Hence
\begin{equation}
\mathcal{I}:\qquad \mathbf{k}\to-\mathbf{k},\qquad \mathbf{J}\to \mathbf{J},\qquad \varepsilon\to \varepsilon.
\label{eq:inversionRules}
\end{equation}

\subsection{Symmetry of individual Hamiltonian terms}

We determine the point groups of the relevant combinations of terms in the total Hamiltonian, Eq.~\ref{eq:HTotal}. The results of the symmetry analysis below are summarized in Table \ref{tab:pointGroupSummary} in the main text.

\bigskip\noindent\emph{$H_{\mathrm{Kane}}$}---
The Kane Hamiltonian (without explicit inversion-asymmetry invariants) is constructed from cubic invariants that are even in $\mathbf{k}$. Under inversion, $H_{\mathrm{Kane}}$ transforms as
\begin{equation}
\mathcal{I}H_{\mathrm{Kane}}(\mathbf{k})\mathcal{I}^{-1}=H_{\mathrm{Kane}}(\mathbf{-k}),
\label{eq:KaneEvenInK}
\end{equation}
so inversion is a symmetry of this truncated model. Together with invariance under the cubic rotations, this combines to the point group $O_h$.

\bigskip\noindent\emph{$H_{\mathrm{Kane}}+H_{\mathrm{Pikus\text{--}Bir}}$}---
For biaxial strain, the strain environment is tetragonal. This can be shown explicitly by checking that $\varepsilon$ is invariant under a $C_{4z}$ rotation:
\begin{equation}
R_{C_{4z}}=
\begin{pmatrix}
0 & -1 & 0\\
1 & 0 & 0\\
0 & 0 & 1
\end{pmatrix},
\qquad
R_{C_{4z}}\,\varepsilon\,R_{C_{4z}}^{T}
=\varepsilon,
\label{eq:C4LeavesStrainInvariant}
\end{equation}
where the last equality follows from $\varepsilon_{xx}=\varepsilon_{yy}$.
Since $H_{\mathrm{Pikus\text{--}Bir}}$ depends only on the inversion-even tensor $\varepsilon$ and angular-momentum operators, it does not introduce inversion-odd ($\mathbf{k}\to-\mathbf{k}$). Therefore the combined $H_{\mathrm{Kane}}+H_{\mathrm{Pikus\text{--}Bir}}$ is tetragonal and inversion-symmetric. Thus, the point group is $D_{4h}$.

\bigskip\noindent\emph{$H_{\mathrm{Kane}}+H_{\mathrm{BIA}}$}---
The bulk inversion asymmetry term contains zincblende invariants that are odd in $\mathbf{k}$ (see Appendix section~\ref{biadef} for the explicit $\mathbf{k}$-dependent blocks). 
At the level of invariants, inversion is broken, because under Eq.~\eqref{eq:inversionRules}
\begin{equation}
\mathcal{I}H_\mathrm{BIA}(k) \mathcal{I}^{-1} = H_{\mathrm{BIA}}(-\mathbf{k})=-H_{\mathrm{BIA}}(\mathbf{k}),
\label{eq:BIAOddInK}
\end{equation}
so $\mathcal I$ is not a symmetry once $H_{\mathrm{BIA}}$ is included. The remaining proper cubic symmetry is that of zincblende. Thus, the point group is $T_d$.

\bigskip\noindent\emph{$H_{\mathrm{Kane}}+H_{\mathrm{Pikus\text{--}Bir}}+H_{\mathrm{BIA}}$}--- 
Starting from $H_{\mathrm{Kane}}+H_{\mathrm{Pikus\text{--}Bir}}$, inversion is removed once $H_{\mathrm{BIA}}$ is included [Eq.~\eqref{eq:BIAOddInK}], while the tetragonal axis set by $\varepsilon_{xx}=\varepsilon_{yy}$ remains. The resulting non-centrosymmetric tetragonal group is $D_{2d}$.

\bigskip\noindent\emph{$H_{\mathrm{Kane}}+H_{\mathrm{Pikus\text{--}Bir}}+H_{C_4}$}---
We first consider how the $C_4$ strain terms [Eqs.~\eqref{eq:hc4}--\eqref{defC487}] transform under inversion
%
% The $C_4$ strain term is defined by 
% \begin{equation}
% H^{8v,8v}_{C4}
% =
% C_4\Big[(\varepsilon_{yy}-\varepsilon_{zz})k_x J_x
% +(\varepsilon_{zz}-\varepsilon_{xx})k_y J_y
% +(\varepsilon_{xx}-\varepsilon_{yy})k_z J_z\Big],
% \label{eq:HC4def}
% \end{equation}
% and similarly $H^{8v,7v}_{C4}\propto \sum_i(\cdots)\varepsilon\,k_iU_i$. 
%
%\textit{Inversion parity of $H_{C_4}$
Using Eq.~\eqref{eq:inversionRules}, we have $\varepsilon\to\varepsilon$, $k_i\to -k_i$, and $J_i\to J_i$ (axial vector).  Hence  products of the form $k_iJ_i$ changes sign,
\begin{equation}
\mathcal I\,(k_iJ_i)\,\mathcal I^{-1}=(-k_i)(J_i)=-(k_iJ_i),
\label{eq:kJInversion}
\end{equation}
and therefore
\begin{equation}
\mathcal I\,H^{8v,8v}_{C_4}(\mathbf{k},\varepsilon)\,\mathcal I^{-1}
=H^{8v,8v}_{C_4}(-\mathbf{k},\varepsilon)=-H_{C_4}(\mathbf{k},\varepsilon).
\label{eq:HC4OddInversion}
\end{equation}
The term $H^{8v,7v}_{C_4}$ transforms analogously.
Thus $H_{C_4}$ is \emph{odd under inversion} and breaks inversion symmetry when added to a centrosymmetric parent Hamiltonian.

For biaxial strain $\varepsilon_{xx}=\varepsilon_{yy}=\varepsilon_\parallel$, the third term in Eq.~\eqref{defC488} vanishes, and one may write
\begin{equation}
H^{8v,8v}_{C_4}\propto (\varepsilon_\parallel-\varepsilon_\perp)\,(k_xJ_x-k_yJ_y).
\label{eq:HC4biaxialForm}
\end{equation}

Adding $H_{C_4}$ to $H_{\mathrm{Kane}}+H_{\mathrm{Pikus\text{--}Bir}}$, we thus break inversion symmetr yby the inversion-odd nature of $H_{C_4}$ [Eq.~\eqref{eq:HC4OddInversion}]. To verify that the remaining symmetry is compatible with $D_{2d}$ (and not lower), we explicitly check invariance of the biaxial form Eq.~\eqref{eq:HC4biaxialForm} under two characteristic operations of $D_{2d}$:

(a) Diagonal mirror $\sigma_d$ (reflection in the plane $x=y$). This swaps $x\leftrightarrow y$ and is improper, so $\det R_{\sigma_d}=-1$.
Thus $k_x\leftrightarrow k_y$ and, by Eq.~\eqref{eq:JTransform}, $J_x\mapsto -J_y$ and $J_y\mapsto -J_x$.
Therefore
\begin{equation}
k_xJ_x-k_yJ_y \xrightarrow{\sigma_d} k_y(-J_y)-k_x(-J_x)=k_xJ_x-k_yJ_y,
\label{eq:sigma_d_invariance}
\end{equation}
so $\sigma_d$ is a symmetry.

(b) Improper fourfold rotation $S_4=\mathcal I C_{4z}$. Under $C_{4z}$, one has
$k_x\to k_y$, $k_y\to -k_x$, and $J_x\to J_y$, $J_y\to -J_x$ (proper rotation, $\det=+1$), which gives
\begin{equation}
k_xJ_x-k_yJ_y \xrightarrow{C_{4z}} k_yJ_y-(-k_x)(-J_x)=-(k_xJ_x-k_yJ_y).
\label{eq:C4_flips}
\end{equation}
Under inversion $\mathcal I$, Eq.~\eqref{eq:inversionRules} implies $k_xJ_x-k_yJ_y\to-(k_xJ_x-k_yJ_y)$.
Combining both actions, $S_4=\mathcal IC_{4z}$ leaves Eq.~\eqref{eq:HC4biaxialForm} invariant.
Hence $H_{\mathrm{Kane}}+H_{\mathrm{Pikus\text{--}Bir}}+H_{C_4}$ is tetragonal and non-centrosymmetric with $\sigma_d$ and $S_4$ symmetries. Thus the symmetry is $D_{2d}$.

\bigskip\noindent\emph{$H_{\mathrm{Kane}}+H_{\mathrm{Pikus\text{--}Bir}}+H_{\mathrm{BIA}}+H_{C_4}$}---
From $H_{\mathrm{Kane}}+H_{\mathrm{Pikus\text{--}Bir}}+H_{\mathrm{BIA}}$, the symmetry is already $D_{2d}$. Adding $H_{C_4}$ does not further reduce the point group because $H_{C_4}$ is compatible with the same $D_{2d}$ generators explicitly checked in $H_{\mathrm{Kane}}+H_{\mathrm{Pikus\text{--}Bir}}+H_{C_4}$. The point group of this combination, equal to $H_\mathrm{Total}$, is thus $D_{2d}$.

\section{Symmetries at the high-symmetry lines $\Gamma$--$X$, $\Gamma$--$K$, and $\Gamma$--$L$}
\label{app:high-symmetry-lines}

We now analyze in detail how the symmetry of the crystal momentum $\mathbf{k}$ along
high-symmetry directions constrains the form of the effective Hamiltonian and,
consequently, the allowed sub-band splittings away from $\Gamma$.
The analysis is carried out entirely at the invariant level, without reference to
explicit matrix representations.

\subsection{General invariant formulation for a Kramers pair}\label{geninv}

Consider a pair of bands that are degenerate at $\Gamma$ and remain well isolated
from other bands along a given high-symmetry line. Restricting the full Hamiltonian
to this two-dimensional subspace yields the most general Hermitian form\cite{PhysRevB.76.045302}
\begin{equation}
H_{\mathrm{eff}}(\mathbf{k})
=
\varepsilon_0(\mathbf{k})\,\mathbb{1}
+
\sum_{i=x,y,z} d_i(\mathbf{k})\,\sigma_i,
\label{eq:Heff_general_detailed}
\end{equation}
where $\sigma_i$ are Pauli matrices.

Demanding that $H_\mathrm{eff}$ be invariant under time reversal,
\begin{equation}
H_{\mathrm{eff}}(\mathbf{k})
=
\sigma_y H_{\mathrm{eff}}^{\ast}(-\mathbf{k}) \sigma_y^{-1},
\label{eq:TRconstraint_detailed}
\end{equation}
implies that 
\begin{equation}
\varepsilon_0(\mathbf{k})=\varepsilon_0(-\mathbf{k}),
\qquad
\mathbf{d}(\mathbf{k})=-\mathbf{d}(-\mathbf{k}).
\label{eq:d_TRodd}
\end{equation}
Thus any splitting $\Delta E(\mathbf{k})=2|\mathbf{d}(\mathbf{k})|$ is necessarily even in $\mathbf{k}$ and vanishes at $\Gamma$.

Now let $g$ be an element of the \emph{little group} of $\mathbf{k}$, i.e.\ a point-group
operation such that $R_g\mathbf{k}=\mathbf{k}$. In the two-dimensional subspace, $g$
is represented by a unitary matrix $U_g$.
Symmetry requires
\begin{equation}
U_g H_{\mathrm{eff}}(\mathbf{k}) U_g^{-1}
=
H_{\mathrm{eff}}(\mathbf{k}).
\label{eq:little_group_constraint}
\end{equation}
Using $U_g \sigma_i U_g^{-1} = \sum_j R^{(s)}_{ij}(g)\sigma_j$, where
$R^{(s)}(g)$ is a $3\times3$ rotation or reflection matrix acting in the pseudospin
space, Eq.~\eqref{eq:little_group_constraint} yields
\begin{equation}
\mathbf{d}(\mathbf{k}) = R^{(s)}(g)\,\mathbf{d}(\mathbf{k})
\quad \text{for all } g \text{ in the little group.}
\label{eq:d_fixed_point}
\end{equation}
The allowed splitting directions are therefore given by the intersection of the
fixed subspaces of all $R^{(s)}(g)$.

\subsection{Little groups at high symmetry lines}

\emph{High symmetry line $\Gamma$--$X$: little group $C_{2v}$}---
Along $\Gamma$--$X$ we have $\mathbf{k}=(k,0,0)$.
The little group of this wavevector is $C_{2v}$\cite{winkler2003spin,PhysRev.100.573,koster63}, generated by:
(i) a twofold rotation $C_{2x}$ about the $x$ axis, and
(ii) two mirror planes $\sigma_{xz}$ and $\sigma_{xy}$.

We now apply each generator to $\mathbf{d}(\mathbf{k})$.
%
%\paragraph{Twofold rotation $C_{2x}$.}
Under $C_{2x}$,
$
(k_x,k_y,k_z)\mapsto(k_x,-k_y,-k_z),
$
and the pseudospin transforms as an axial vector,
$
(d_x,d_y,d_z)\mapsto(d_x,-d_y,-d_z).
$
Equation~\eqref{eq:d_fixed_point} therefore requires
\begin{equation}
d_y(\mathbf{k})=d_z(\mathbf{k})=0.
\label{eq:C2x_constraint}
\end{equation}
%
%\paragraph{Mirror $\sigma_{xy}$.}
Under reflection in the $xy$ plane,
$
(k_x,k_y,k_z)\mapsto(k_x,k_y,-k_z),
$
and axial vectors transform with an extra sign,
$
(d_x,d_y,d_z)\mapsto(-d_x,-d_y,d_z).
$
Invariance then requires
\begin{equation}
d_x(\mathbf{k})=d_y(\mathbf{k})=0.
\label{eq:sigma_xy_constraint}
\end{equation}
Combining Eqs.~\eqref{eq:C2x_constraint} and \eqref{eq:sigma_xy_constraint}, we obtain
\begin{equation}
\mathbf{d}(\mathbf{k})=\mathbf{0}
\quad \text{to linear order in } \mathbf{k}.
\label{eq:GX_no_linear_split}
\end{equation}
Thus, symmetry forces the linear-in-$k$ pseudospin splitting to vanish along
$\Gamma$--$X$.
This explains why inversion-asymmetry-induced splitting from $H_{\mathrm{BIA}}$ is
strongly suppressed along this direction.

By contrast, the strain-induced $C_4$ term reduces along $\Gamma$--$X$ to
\begin{equation}
H_{C_4}\propto k_x J_x,
\label{eq:HC4_GX_final}
\end{equation}
which acts \emph{within} the valence-band manifold and produces a finite splitting
through band mixing rather than a direct pseudospin field $\mathbf{d}$.
%This is why $H_{C_4}$ dominates the observed splitting on $\Gamma$--$X$.

\bigskip\noindent\emph{High-symmetry line $\Gamma$--$K$: little group $C_s$}---
Along $\Gamma$--$K$ in the $k_z=0$ plane, we parametrize the momentum vectors as $\mathbf{k}=(\lambda,\lambda,0)$.
The little group is $C_s$\cite{winkler2003spin,PhysRev.100.573,koster63}, generated by a single mirror plane $\sigma_d$ containing
the $\Gamma$--$K$ direction.

Under $\sigma_d$, momentum transforms as
$
(k_x,k_y,k_z)\mapsto(k_y,k_x,k_z),
$
while the axial pseudospin transforms as
$
(d_x,d_y,d_z)\mapsto(-d_y,-d_x,-d_z).
$
Equation~\eqref{eq:d_fixed_point} then yields
\begin{equation}
d_x(\mathbf{k})=-d_y(\mathbf{k}),\qquad d_z(\mathbf{k})=0.
\label{eq:Cs_constraint}
\end{equation}
Unlike the $C_{2v}$ case, this constraint allows a \emph{one-dimensional subspace}
of nonzero $\mathbf{d}$ vectors.
Consequently, linear-in-$k$ splitting is symmetry-allowed along $\Gamma$--$K$.

The $C_4$ strain term reduces to
\begin{equation}
H_{C_4}\propto \lambda\,(J_x-J_y),
\label{eq:HC4_GK_final}
\end{equation}
which contributes weakly to the splitting of the relevant hybridized states.
In contrast, $H_{\mathrm{BIA}}$ generates a pseudospin field consistent with
Eq.~\eqref{eq:Cs_constraint}, leading to an enhanced splitting along $\Gamma$--$K$.

\bigskip\noindent\emph{High-symmetry line $\Gamma$--$L$: little group $C_{3v}$}---
Along $\Gamma$--$L$ we consider the (111) direction,
\begin{equation}
\mathbf{k}=(\lambda,\lambda,\lambda),
\qquad
\lambda=\frac{k}{\sqrt{3}}.
\label{eq:k_GL}
\end{equation}
The little group is $C_{3v}$\cite{winkler2003spin,PhysRev.100.573,koster63}, generated by a threefold rotation $C_{3}$ about the
$(111)$ axis and three vertical mirror planes.

Under $C_{3}$, the pseudospin transforms as a vector rotated by $120^\circ$ about
$(1,1,1)$. The only vectors left invariant by such a rotation are those parallel to
the rotation axis. Hence Eq.~\eqref{eq:d_fixed_point} implies
\begin{equation}
\mathbf{d}(\mathbf{k}) \parallel (1,1,1).
\label{eq:C3v_constraint}
\end{equation}
This one-dimensional invariant subspace allows a finite linear-in-$k$ splitting
consistent with $C_{3v}$ symmetry.

The $C_4$ strain term contributes negligibly along this direction due to the
cancellation of strain differences in the biaxial geometry, whereas
$H_{\mathrm{BIA}}$ naturally produces a pseudospin field aligned with the $(111)$
axis. As a result, the splitting along $\Gamma$--$L$ is large and dominated by BIA.

\section{The topological phase diagram of HgTe as a function of strain}\label{stopo}
Based on our study of the evolution of the Weyl semimetal state with strain, we construct a topological phase diagram of \ce{HgTe} as a function of strain (Fig. \ref{fig.S1}(b)). Our results, obtained from fitting $H_\mathrm{Total}$ to the band structure calculated using DFT, are consistent with those predicted in prior work \cite{ruan2016symmetry,chen2019strain}, as a result of which we obtain a similar topological phase diagram, with the exception of a tilted type-1 Weyl semimetal instead of an ideal type-1 Weyl semimetal state at large compressive strains. This demonstrates that the inclusion of the $C_4$ strain terms in our model does not change the topology of \ce{HgTe}.

% Table belongs to appendix F (Position and chirality of Weyl nodes
\begin{table}[t]
\centering
\begin{tabular}{|l|l|l|l|}
\hline
$k_x$ ($\mathrm{{\mathring{A}}}^{-1}$) & $k_y$ ($\mathrm{{\mathring{A}}}^{-1}$) & $k_z$ ($\mathrm{{\mathring{A}}}^{-1}$) & Chirality\\ \hline
0.001438 & 0.0 & 0.017871 & +1\\ \hline
0.001438 & 0.0 & -0.017871 & +1\\ \hline
-0.001438 & 0.0 & 0.017871 & +1\\ \hline
-0.001438 & 0.0 & -0.017871 & +1\\ \hline
0.0 & 0.001440 & 0.017881 & -1\\ \hline
0.0 & 0.001440 & -0.017881 & -1\\ \hline
0.0 & -0.001440 & 0.017881 & -1\\ \hline
0.0 & -0.001440 & -0.017881 & -1\\ \hline
\end{tabular}
\caption{\label{table3} The Weyl points calculated at $\epsilon_{xx}=-0.5\%$ and their associated chirality.}
\end{table}

\section{Characterization of Weyl nodes}

\subsection{Position and chirality}

%%% Table, see above

We calculate the chirality of Weyl nodes obtained at $\epsilon_{xx}=-0.5\%$ strain using the methodology described in Ref.~\cite{ruan2016symmetry}. 
Our results are consistent with these prior studies and have been provided in Table \ref{table3}.

\subsection{Quantification of Weyl cone tilt}\label{quant}

To characterize the nature of the Weyl nodes, we perform a local expansion of the two crossing bands around the Weyl point. Near a Weyl node located at $\mathbf{k}_W$, the low-energy Hamiltonian can be written in the standard linearized form, where energy is measured relative to the energy $E(\mathbf{k_{W}})$ of the Weyl node \cite{RevModPhys.90.015001,soluyanov2015type},
\begin{equation}
H(\mathbf{q}) = \mathbf{w} \cdot \mathbf{q}\,\sigma_0 + \sum_{i,j} v_{ij} q_i \sigma_j,\label{eq12}
\end{equation}
where $\mathbf{q} = \mathbf{k} - \mathbf{k}_W$, $\mathbf{w}$ is the tilt vector, and $v_{ij}$ describes the anisotropic Weyl cone velocities, where $(i,j) = (x,y,z)$.

The eigenergies of Eq. \eqref{eq12}, are given by :
\begin{equation}
E_\pm(\mathbf{q}) = S(\mathbf{q}) \pm D(\mathbf{q})
\end{equation}
with
\begin{equation}
S(\mathbf{q}) = \mathbf{w} \cdot \mathbf{q}, \quad
D(\mathbf{q})^2 = \mathbf{q}^T G \mathbf{q},
\end{equation}
where $G = v v^T$ is a positive-definite metric matrix describing the anisotropic dispersion.
We determine $S(\mathbf{q})=(E_+(\mathbf{q}) + E_-(\mathbf{q})) / 2$ and $D(\mathbf{q}) = (E_+(\mathbf{q}) - E_-(\mathbf{q})) /2$ from a least squares fit  to the numerically obtained dispersions $E_{+}(\mathbf{q})$ and $E_{-}(\mathbf{q})$ over a three-dimensional momentum grid in the vicinity of the Weyl point
The strength of the tilt is then quantified by the dimensionless parameter\cite{soluyanov2015type}
\begin{equation}
\kappa_{\max} = \sqrt{\mathbf{w}^T G^{-1} \mathbf{w}}.
\end{equation}
\\
A Weyl node is classified as type-I for $\kappa_{\max} < 1$ and type-II for $\kappa_{\max} > 1$. For the Weyl node obtained from $H_{Total}$ in Table \ref{table2} at $\epsilon_{xx} = -0.5\%$, we obtain $\kappa_{\max} = 0.846$, indicating a finite tilt while remaining in the type-I regime.

\end{appendix}
\bibliography{bibliography.bib}


% Provide your bibliography here. You have two options:

%%% FIRST OPTION
% Write your entries here directly, following the example below, including:
% Author(s), Title, Journal Ref. with year in parentheses at the end, followed by the DOI number.

%\begin{thebibliography}{99}
%\bibitem{1931_Bethe_ZP_71} H. A. Bethe, {\it Zur Theorie der Metalle. i. Eigenwerte und Eigenfunktionen der linearen Atomkette}, Zeit. f{\"u}r Phys. {\bf 71}, 205 (1931), \doi{10.1007\%2FBF01341708}.
%\bibitem{arXiv:1108.2700} P. Ginsparg, {\it It was twenty years ago today... }, \url{http://arxiv.org/abs/1108.2700}.
%\end{thebibliography}

% Use your bibtex library, formatted by the SciPost style file.
%\bibliography{SciPost_Example_BiBTeX_File.bib}

%%%%%%%%%% END TODO: BIBLIOGRAPHY

\end{document}